\documentclass[12pt,a4paper] {article}
\usepackage{epsfig}
\usepackage{rotating}
\usepackage{cite}
\begin{document}
\newcommand{\omclb}      {\mbox{${\mathrm {1-CL_b}}$}}
\newcommand{\clb}        {\mbox{${\mathrm {CL_b}}$}}
\newcommand{\cls}        {\mbox{${\mathrm {CL_s}}$}}
\newcommand{\clsb}       {\mbox{${\mathrm {CL_{s+b}}}$}}
\newcommand{\Z}{\mbox{$\mathrm{Z}$}}
\newcommand{\Zo}{\mbox{$\mathrm{Z^0}$}}
\newcommand{\W}{\mbox{$\mathrm{W}$}}
\newcommand{\bZo}{{\bf \mbox{$\mathrm{Z}$}}}
\newcommand{\Zg}{\mbox{$\mathrm{Z}^{0}\gamma$}}
\newcommand{\ZZ}{\mbox{$\mathrm{Z}^{0}\mathrm{Z}^{0}$}}
\newcommand{\WW}{\mbox{$\mathrm{W}\mathrm{W}$}}
\newcommand{\Zs}{\mbox{$\mathrm{Z}^{*}$}}
\newcommand{\h}{\mbox{$\mathrm{h}^{0}$}}
\newcommand{\Ho}{\mbox{$\mathrm{H}$}}
\newcommand{\ho}{\mbox{$\mathrm{h}$}}
\newcommand{\Hp}{\mbox{$\mathrm{H}^{+}$}}
\newcommand{\Hm}{\mbox{$\mathrm{H}^{-}$}}
\newcommand{\Hsm}{\mbox{$\mathrm{H}^{0}_{SM}$}}
\newcommand{\A}{\mbox{$\mathrm{A}^{0}$}}
\newcommand{\Hpm}{\mbox{$\mathrm{H}^{\pm}$}}
\newcommand{\X}{\mbox{${\tilde{\chi}^0}$}}
\newcommand{\ko}{\mbox{${\tilde{\chi}^0}$}}
\newcommand{\ee}{\mbox{$\mathrm{e}^{+}\mathrm{e}^{-}$}}
\newcommand{\bee}{\mbox{$\boldmath {\mathrm{e}^{+}\mathrm{e}^{-}} $}}
\newcommand{\mm}{\mbox{$\mu^{+}\mu^{-}$}}
\newcommand{\nn}{\mbox{$\nu \bar{\nu}$}}
\newcommand{\qq}{\mbox{$\mathrm{q} \bar{\mathrm{q}}$}}
\newcommand{\pb}{\mbox{$\mathrm{pb}^{-1}$}}
\newcommand{\ra}{\mbox{$\rightarrow$}}
\newcommand{\br}{\mbox{$\boldmath {\rightarrow}$}}
\newcommand{\tautau}{\mbox{$\tau^{+}\tau^{-}$}}
\newcommand{\ga}{\mbox{$\gamma$}}
\newcommand{\gamgam}{\mbox{$\gamma\gamma$}}
\newcommand{\tp}{\mbox{$\tau^+$}}
\newcommand{\tm}{\mbox{$\tau^-$}}
\newcommand{\tpm}{\mbox{$\tau^{\pm}$}}
\newcommand{\uu}{\mbox{$\mathrm{u} \bar{\mathrm{u}}$}}
\newcommand{\dd}{\mbox{$\mathrm{d} \bar{\mathrm{d}}$}}
\newcommand{\bb}{\mbox{$\mathrm{b} \bar{\mathrm{b}}$}}
\newcommand{\cc}{\mbox{$\mathrm{c} \bar{\mathrm{c}}$}}
\newcommand{\mumu}{\mbox{$\mu^+\mu^-$}}
\newcommand{\csbar}{\mbox{$\mathrm{c} \bar{\mathrm{s}}$}}
\newcommand{\cbars}{\mbox{$\bar{\mathrm{c}}\mathrm{s}$}}
\newcommand{\nunu}{\mbox{$\nu \bar{\nu}$}}
\newcommand{\nubar}{\mbox{$\bar{\nu}$}}
\newcommand{\mQ}{\mbox{$m_{\mathrm{Q}}$}}
\newcommand{\mZ}{\mbox{$m_{\mathrm{Z}}$}}
\newcommand{\mH}{\mbox{$m_{\mathrm{H}}$}}
\newcommand{\mHrec}{\mbox{$m_{\mathrm{H}}^{\mathrm{rec}}$}}
\newcommand{\mHp}{\mbox{$m_{\mathrm{H}^+}$}}
\newcommand{\mh}{\mbox{$m_{\mathrm{h}}$}}
\newcommand{\mA}{\mbox{$m_{\mathrm{A}}$}}
\newcommand{\mHpm}{\mbox{$m_{\mathrm{H}^{\pm}}$}}
\newcommand{\mHsm}{\mbox{$m_{\mathrm{H}^0_{SM}}$}}
\newcommand{\mW}{\mbox{$m_{\mathrm{W}^{\pm}}$}}
\newcommand{\mt}{\mbox{$m_{\mathrm{t}}$}}
\newcommand{\mb}{\mbox{$m_{\mathrm{b}}$}}
\newcommand{\lpm}{\mbox{$\ell ^+ \ell^-$}}
\newcommand{\G}{\mbox{$\mathrm{GeV}$}}
\newcommand{\Gc}{\mbox{${\rm GeV}/c$}}
\newcommand{\Gcs}{\mbox{${\rm GeV}/c^2$}}
\newcommand{\Mcs}{\mbox{${\rm MeV}/c^2$}}
\newcommand{\sba}{\mbox{$\sin ^2 (\beta -\alpha)$}}
\newcommand{\cba}{\mbox{$\cos ^2 (\beta -\alpha)$}}
\newcommand{\tanb}{\mbox{$\tan \beta$}}
\newcommand{\sqrts}{\mbox{$\sqrt {s}$}}
\newcommand{\sqrtsp}{\mbox{$\sqrt {s'}$}}
\newcommand{\msusy}{\mbox{$M_{\rm SUSY}$}}
\newcommand{\mg}{\mbox{$m_{\tilde{\rm g}}$}}
\begin{titlepage}
\begin{center}
\vspace{-0.5cm}
{\Large EUROPEAN ORGANIZATION FOR NUCLEAR RESEARCH}
\end{center}
\bigskip
\begin{flushright}
CERN-EP/2003-011\\
13 March 2003 \\
\end{flushright}
\bigskip
\begin{center}{\Large \bf Search for the Standard Model Higgs Boson at LEP}
\end{center}
\begin{center}
      {\Large  ALEPH, DELPHI, L3 and OPAL Collaborations\\
      The LEP Working Group for Higgs Boson Searches\footnote{The authors 
      are listed in Refs.~\cite{a_new,d_new,l_new,o_new,lhwg}.}} 
\end{center}
\bigskip
\begin{center}{\Large  Abstract}\end{center}
The four LEP collaborations, ALEPH, DELPHI, L3 and OPAL, have collected a total of
2461~pb$^{-1}$ of \ee\ collision data at centre-of-mass energies between 189 and 209 GeV. 
The data are used to search for the Standard Model Higgs boson.
The search results of the four collaborations are combined and examined in a likelihood test
for their consistency with two hypotheses: the background hypothesis and the signal plus background 
hypothesis. The corresponding confidences have been computed as 
functions of the hypothetical Higgs boson mass. 
A lower bound 
of 114.4~\Gcs\ is established, at the 95\% confidence level, on the mass of the Standard Model Higgs boson.
The LEP data are also used to set upper bounds on the HZZ coupling for various assumptions concerning the decay of the Higgs boson.
\bigskip
$\phantom{.....}$\\
$\phantom{.....}$\\
$\phantom{.....}$\\
$\phantom{.....}$\\
$\phantom{.....}$\\
\begin{center}
{\it To be submitted to Physics Letters B}
\end{center}
\end{titlepage}
\section{Introduction}
The Higgs mechanism~\cite{higgs} plays a central role in the unification of the electromagnetic and weak interactions
by providing mass to the \W\ and \Z\ intermediate vector bosons without violating local gauge invariance.
Within the Standard Model~\cite{sm}, the Higgs mechanism is invoked to break the electroweak symmetry; it implies 
the existence of a 
single neutral scalar particle, the Higgs boson. The mass of this particle is not specified, but
indirect experimental limits are obtained from precision measurements of the
electroweak parameters which depend logarithmically on the Higgs boson mass through radiative corrections. 
Currently these measurements predict that the Standard Model Higgs boson mass is $m_{\rm H}= 81^{+52}_{-33}$~\Gcs\
and constrain its value to less 
than 193~\Gcs\ at the 95\% confidence level~\cite{blueband}.    

The data collected by the four LEP collaborations prior to the year 2000 gave no direct indication of the production of the 
Standard Model Higgs boson~\cite{adlo-cernep}
and allowed a lower bound of 107.9~\Gcs\ to be set, at the 95\% confidence level, on the mass.
During the last year of the LEP programme (the year 2000), substantial data samples were collected at centre-of-mass 
energies exceeding 206~GeV, extending the search sensitivity to Higgs boson masses of about 115~\Gcs\ 
through the Higgsstrahlung process \ee\ra~HZ.
In their initial analyses of the full data sets, ALEPH~\cite{a_old} observed an excess of events consistent with the production 
of a Standard Model Higgs boson with a mass of 115~\Gcs; L3~\cite{l_old} and OPAL~\cite{o_old},
while being consistent with the background hypothesis,
slightly favoured the signal plus background hypothesis in this mass region;  
DELPHI~\cite{d_old} reported a slight deficit with respect 
to the background expectation. 
The final results from the four collaborations have now been
published~\cite{a_new,d_new,l_new,o_new}. These are based on final calibrations 
of the detectors and LEP beam energies and, in some cases, on revised analysis procedures. 
In this paper we present the results from a LEP-wide combination based on these new publications. 
The data span the range of centre-of-mass energies (\sqrts~) from 189~GeV to 209~GeV.
The integrated luminosities of the data samples are given in Table~\ref{tab:lumi} for the full range of energies used 
and for the subset with energies larger than 206~GeV.  

We also present upper bounds on the HZZ coupling for non-standard models with various assumptions concerning the 
decay of the Higgs boson. In order to cover the low-mass domain,
the data collected during the LEP1 phase at the \Z\ resonance are combined with LEP2 data.
\begin{table}[htp]
\begin{center}
\begin{tabular}{|c|}
\hline
~~~~~~~~~~~~~~~~~~Integrated luminosities in \pb~~~~~~~~~~~~~~~~~~\\
\hline
\end{tabular}
\begin{tabular}{|l|cccc|c|}
\hline 
                          & ALEPH      & DELPHI     & L3        & OPAL   & LEP   \\
\hline\hline
\sqrts~$\ge 189$ GeV      & 629        & 608        & 627       & 596    & 2461  \\
\sqrts~$\ge 206$ GeV      & 130        & 138        & 139       & 129    & 536   \\
\hline
\end{tabular}
\caption{\small Integrated luminosities of the data samples of the four experiments 
and their sum (LEP). The subsets taken at energies larger than 206~GeV are listed separately. 
\label{tab:lumi}}
\end{center}
\end{table}
\section{Analysis and combination procedure}
At LEP, the Standard Model Higgs boson is expected to be produced mainly in association with the \Z\ boson through the 
Higgsstrahlung process
\ee\ra\Ho\Z~\cite{bjorken}. Small additional contributions are expected 
at the end of the kinematic range of the Higgsstrahlung process from \W\ and \Z\ boson fusion, which produce a Higgs boson and a pair 
of neutrinos or electrons, respectively, in the final state~\cite{fusion}.
The signal processes are simulated using the HZHA generator~\cite{hzha}, which includes the fusion processes and their interference 
with the HZ final states.
For Higgs boson masses which are relevant at LEP, the Standard Model Higgs 
boson is expected to decay mainly into \bb\ quark pairs
(the branching ratio is 74\% for a mass of 115~\Gcs) while decays to $\tau^+\tau^-$, WW$^*$, gg ($\approx$ 7\% each), 
and \cc\ ($\approx$ 4\%) constitute the rest of the decay width. 
The final-state topologies are determined by the decay properties of the Higgs boson and by those of the associated \Z\ boson. 
The searches at LEP encompass the  four-jet final state (\Ho\ra\bb)(Z\ra\qq), the missing
energy final state (\Ho\ra\bb)(Z\ra\nn), the leptonic final state (\Ho\ra\bb)(Z\ra$\ell^+\ell^-$) where $\ell$ denotes an 
electron or a muon, and the tau lepton
final states (\Ho\ra\bb)(Z\ra$\tau^+\tau^-$) and (\Ho\ra$\tau^+\tau^-$)(Z\ra\qq).
 
A preselection is applied by each experiment to reduce some of the main backgrounds, in particular from two-photon processes and from the
radiative return to the \Z\ boson, \ee\ra\Z$\gamma(\gamma)$. 
The remaining background, mainly from fermion pairs and \WW\ or ZZ production, possibly with photon or gluon radiation, is 
further reduced either with the help of more selective cuts or by applying multivariate techniques such as
likelihood analyses and neural networks. 
The identification 
of b-quarks in the decay of the Higgs boson plays an important role in the 
discrimination between signal and background, as does the reconstructed Higgs boson candidate mass.
The detailed implementation of these analyses by the different experiments is described in Refs.~\cite{a_new,d_new,l_new,o_new}
and in earlier references quoted therein. 

The input from the four experiments which is used in the combination procedure is provided channel by channel.  
The word ``channel" designates any subset 
of the data where a Higgs boson search has been carried out. These subsets may correspond 
to specific final-state topologies, to data sets 
collected at different centre-of-mass energies or to the subsets of data collected by different experiments. 
In most channels the input is binned in two 
variables: the reconstructed
Higgs boson mass \mHrec, and a variable $\cal G$ which combines many event features such as b-tagging variables, 
likelihood functions or neural network outputs,
which allow discrimination 
on a statistical basis between the Higgs boson signal 
and the background processes. 

For each given channel and bin in the (\mHrec, $\cal G$) plane, the experiments
provide the number of selected data events, the number of expected
background events, and the number of expected signal events for a set of
hypothetical Higgs boson masses.
The expected signal and background estimates make use of detailed
Monte Carlo simulations carried out by each of the four experiments. These take into account all known 
experimental features such as the centre-of-mass energies and integrated luminosities of the data samples, cross-sections and 
decay branching ratios for the signal and background processes,  
selection efficiencies and experimental resolutions with possible non-Gaussian contributions. Systematic errors with 
their correlations are also evaluated.
Since the simulations are done at fixed centre-of-mass energies and Higgs boson masses,
interpolations are applied (see, for example,~\cite{interpol}) to obtain the rates and
distributions for arbitrary energies and masses. In order to avoid problems which arise
in some bins due to low Monte Carlo statistics, smoothing procedures are applied
(see, for example,~\cite{keys}) which combine
the available information with the information in the neighbouring bins. 
\section{Hypothesis testing}
The observed data configuration in the ($\mHrec ,~\cal G$) plane  is subjected to a likelihood ratio test
of two hypothetical scenarios. In the background scenario 
it is assumed that the data receive contributions from the background processes only, while in the signal plus background
scenario the contributions from a Standard Model Higgs boson of test mass \mH\ are assumed in addition. The expressions for the corresponding 
binned likelihoods ${\cal L}_{\rm b}$ and ${\cal L}_{\rm{s+b}}$
are given in the Appendix. 

In a search experiment, the likelihood ratio
\begin{equation}
         Q= {\cal L}_{\rm{s+b}}/ {\cal L}_{\rm b}
\end{equation}
makes efficient use of the information contained in the event configuration.
For convenience, the logarithmic form $-2\ln Q$ 
is used as the test statistic since this quantity is approximately equal to the difference in $\chi^2$  
when the data configuration is compared to the background hypothesis and to the
signal plus background hypothesis (it becomes exactly equal in the limit of high statistics). 
Furthermore, $-2\ln Q$ can be written as a sum of contributions from the individual observed events (see Eq. (4) in the Appendix).

Figure~\ref{fig:adlo-lnq} shows 
the test statistic $-2\ln Q$ 
as a function of the test mass for the LEP-wide combination.
The expected curves 
are obtained by replacing the observed data configuration by a large number of simulated event 
configurations for the two hypotheses. For the background hypothesis the 68\% and 95\% probability bands are also shown.
There is a broad minimum in the observed $-2\ln Q$ starting at about 115~\Gcs. The negative values in this mass range
indicate that the hypothesis including a Standard Model Higgs boson of such a mass 
is more favoured than the background hypothesis, albeit at low significance.
Note also that the median expectation for the signal plus background hypothesis crosses the observed curve in this mass range.
The fact that the observed curve slightly 
deviates from the background expectation over the whole mass range of the figure can also 
be explained by local upward fluctuations of the background and by
long-range effects due to 
the experimental resolution.
 
In Figure~\ref{fig:a-d-l-o-lnq} 
the likelihood test is applied to subsets of the LEP data from individual experiments and final-state topologies.
A signal-like deviation beyond the 95\% confidence level is only observed in
the ALEPH data. For a given test mass,
the distance between the background expectation and the signal plus background expectation, compared to their spreads,
is a measure of the discriminating power of the corresponding data set. 
These figures thus illustrate the relative power of the subsets and the rapid decrease in discriminating power
as the test mass approaches the kinematic limit of the HZ signal process.
One should note that no individual LEP experiment has the statistical power to distinguish 
between the two hypotheses for a test mass larger than about 114~\Gcs\ at the level of 
two standard deviations (see the intersections of the signal plus background curve with the lower edge of the light-shaded
95\% confidence level bands).
Regarding the final-state topologies, the combined LEP data in the four-jet channel have about the same
discriminating power as all the other final states together. 
The comparison of Figures~\ref{fig:adlo-lnq} and~\ref{fig:a-d-l-o-lnq} illustrates the gain in
sensitivity when the data of the four experiments are combined in all channels.
\subsection{Contributions from single candidates}
\begin{figure}[htb]
\begin{center}
\epsfig{figure=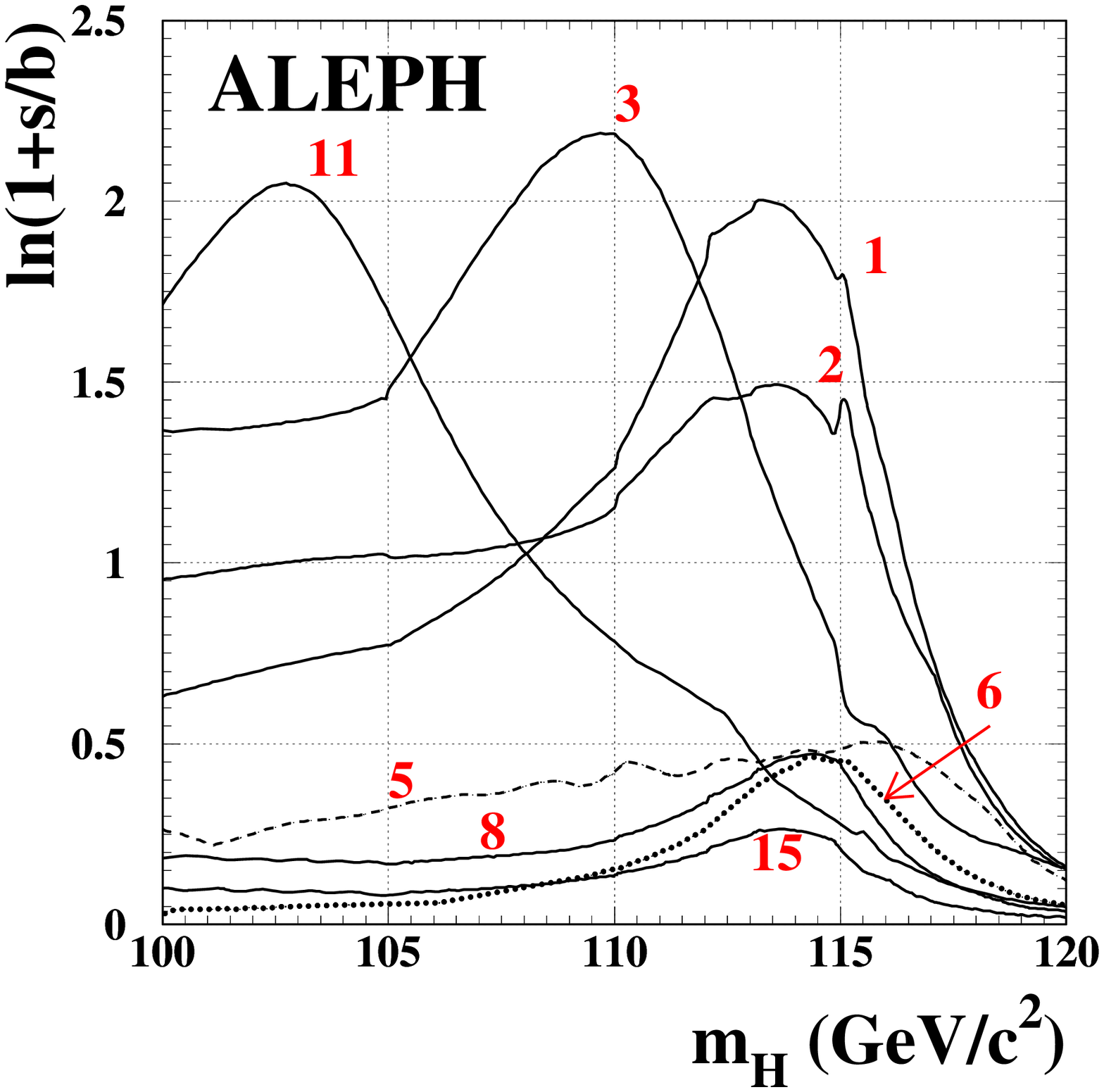,width=0.49\textwidth}
\epsfig{figure=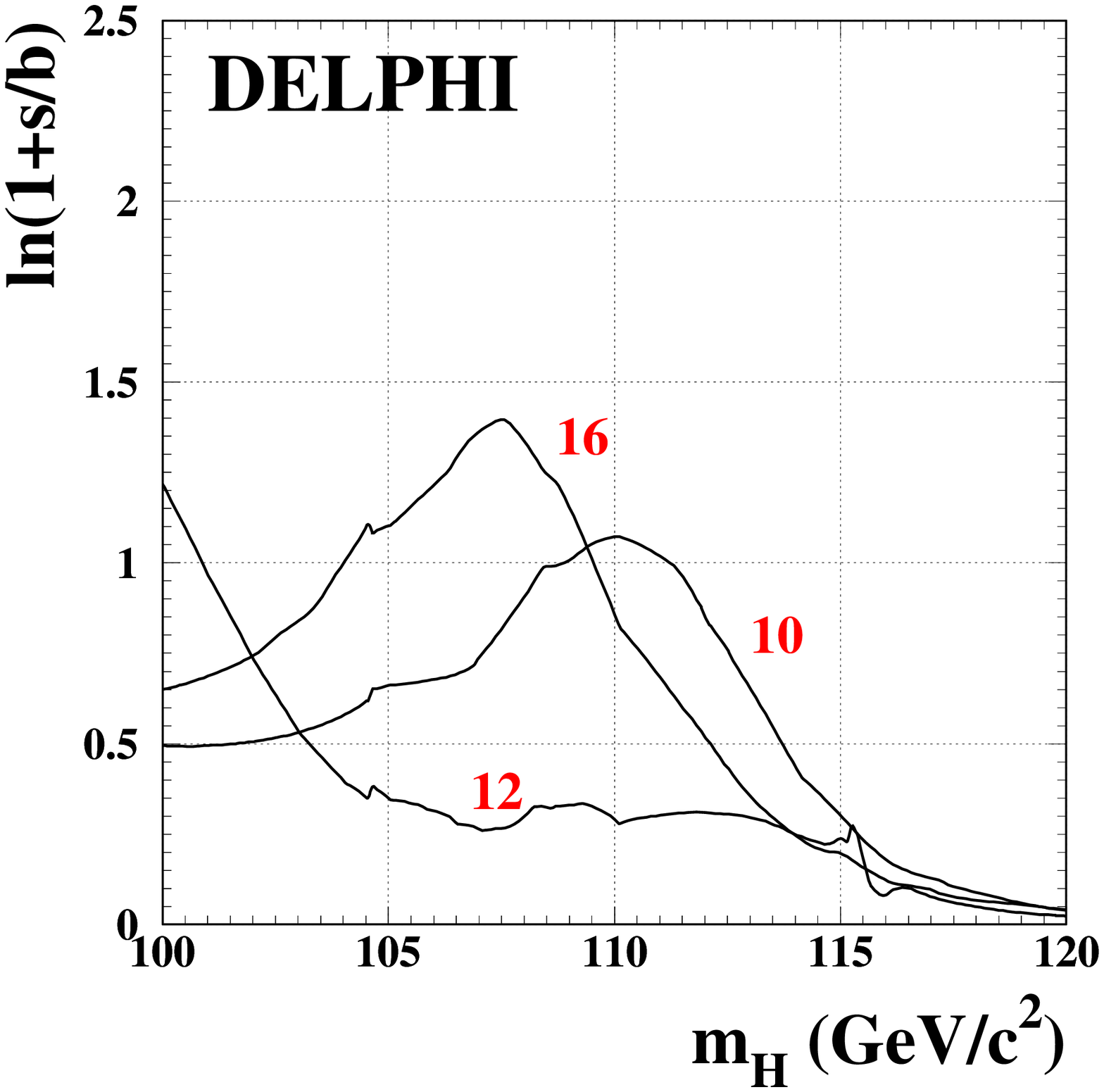,width=0.49\textwidth}\\
\epsfig{figure=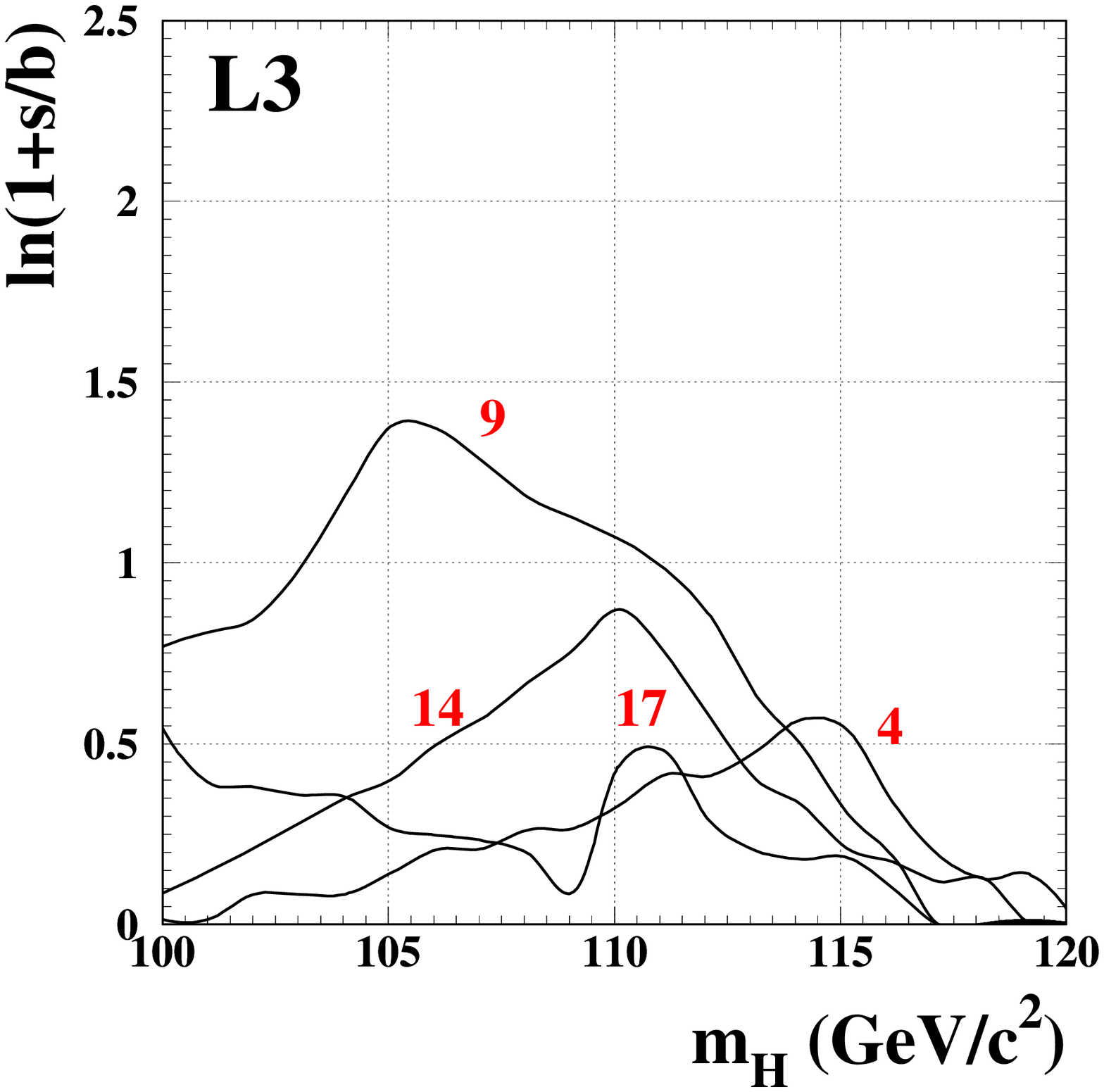,width=0.49\textwidth}
\epsfig{figure=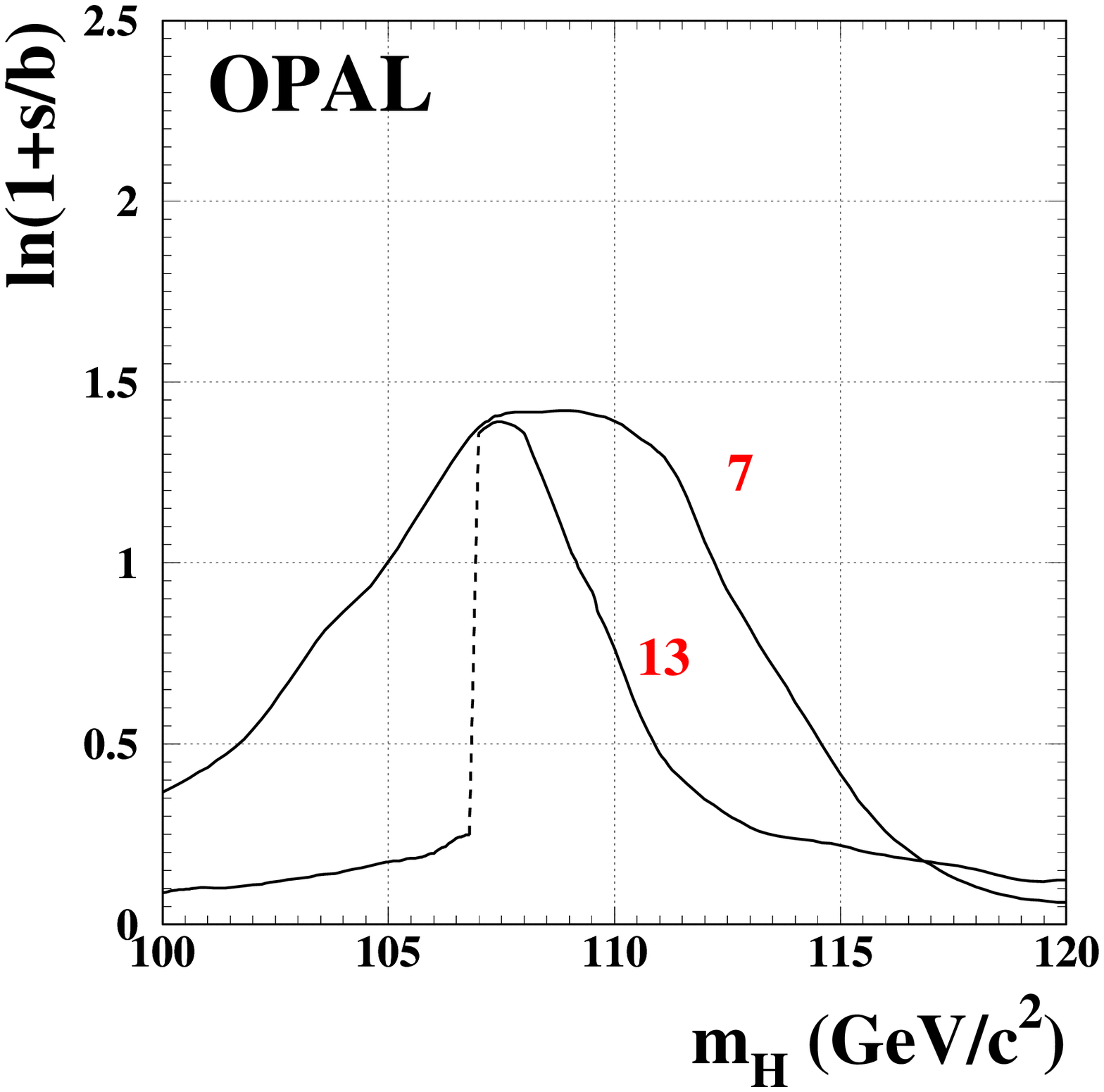,width=0.49\textwidth}
\caption[]{\small 
Evolution of the event weight $\ln (1+s/b)$
with test mass $\mH $ for the events which have the largest contributions to $-2\ln Q$ at $\mH =115$~\Gcs. 
The labels correspond to the candidate numbers in the first column of Table~\ref{tab:event-list}.
The sudden increase in the weight of the OPAL missing energy candidate labelled ``13" at $\mH = 107$~\Gcs\ is 
due to switching from the low-mass to high-mass optimization of the search at that mass. A similar increase is 
observed in the case of the  L3 four-jet candidate labelled ``17" which is due to the test mass 
dependent attribution of the jets to the Z and Higgs bosons.
\label{fig:spaghetti}}
\end{center}
\end{figure}

\begin{figure}[htb]
\begin{center}
\epsfig{figure=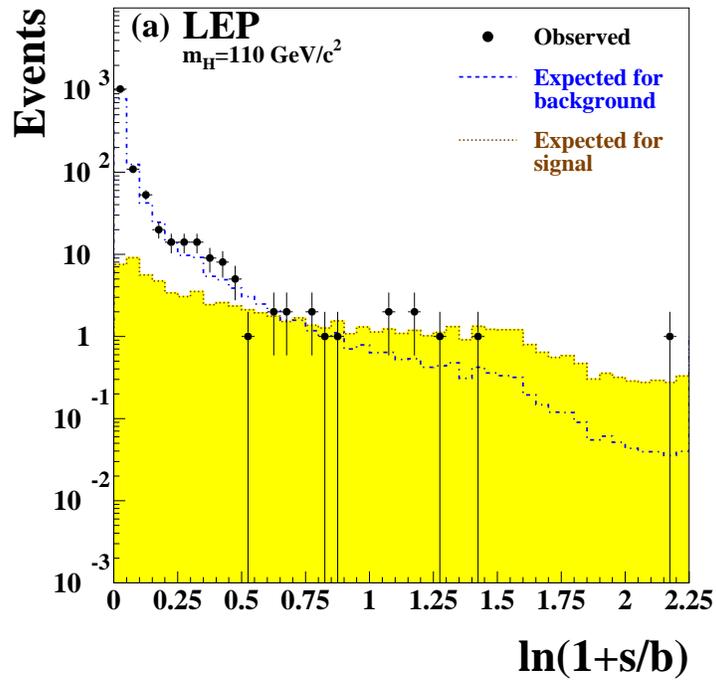,width=0.6\textwidth}\\
\epsfig{figure=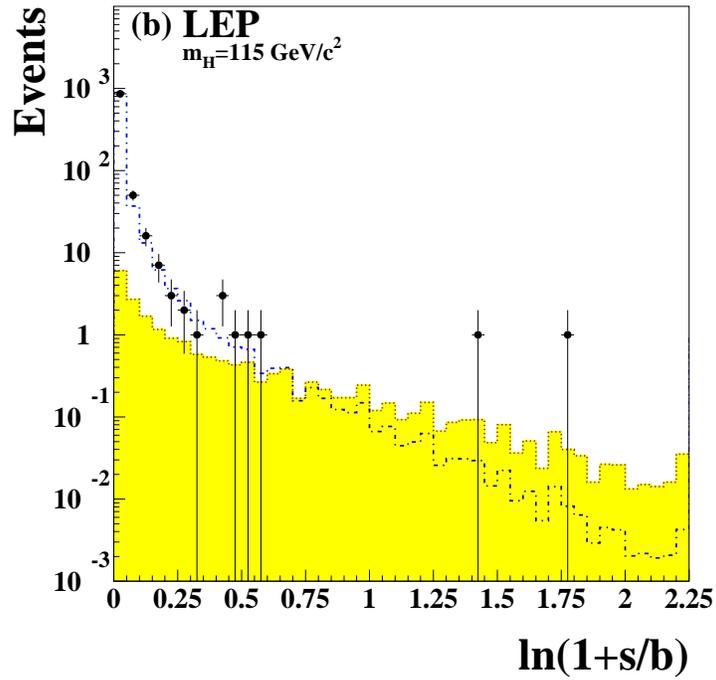,width=0.6\textwidth}\\
\caption[]{\small 
Expected and observed distributions of the event weight $\ln (1+s/b)$ for 
test masses \mH\ of (a) 110 and (b) 115~\Gcs.
Dashed line histograms: expected distributions for the background; shaded histograms: expected distributions for the
signal; points with error bars: selected data.
\label{fig:adlo-weights}}
\end{center}
\end{figure}

The contribution to the test statistic $-2\ln Q$  from an individual candidate event
can be evaluated using the binned likelihood functions that appear in
the Appendix.
We refer to this contribution as the event weight which, in simplified notation, can be
written as $\ln (1 + s/b)$, where $s$ and $b$ refer to the signal and
background estimates in the bins of $(\mHrec,~{\cal G})$ where the events are reconstructed. 
The candidates which have the highest weight for a test mass of 115~\Gcs, chosen throughout this paper for the 
purpose of illustration, 
are listed in Table~\ref{tab:event-list}. 
\begin{table}[t]
\begin{center}
\begin{tabular}{|c|lcl|cc|}
\hline
   & Experiment&  $E_{cm}$(GeV) &  Final state &   $\mHrec$ (\Gcs) &  $\ln (1+s/b)$ \\
   &           &                &  topology    &                            &  at 115 \Gcs\    \\
\hline\hline
1&  ALEPH  &  206.6  &  Four-jet         &   114.1             & 1.76           \\
2&  ALEPH  &  206.6  &  Four-jet         &   114.4             & 1.44           \\
3&  ALEPH  &  206.4  &  Four-jet         &   109.9             & 0.59           \\
4&  L3     &  206.4  &  Missing energy   &   115.0             & 0.53           \\
5&  ALEPH  &  205.1  &  Leptonic         &   117.3             & 0.49           \\
6&  ALEPH  &  208.0  &  Tau              &   115.2             & 0.45           \\
7&  OPAL   &  206.4  &  Four-jet         &   111.2             & 0.43           \\
8&  ALEPH  &  206.4  &  Four-jet         &   114.4             & 0.41           \\
9&  L3     &  206.4  &  Four-jet         &   108.3             & 0.30           \\
10& DELPHI &  206.6  &  Four-jet         &   110.7             & 0.28           \\
11& ALEPH  &  207.4  &  Four-jet         &   102.8             & 0.27           \\
12& DELPHI &  206.6  &  Four-jet         & $\phantom{0}$97.4   & 0.23           \\ 
13& OPAL   &  201.5  &  Missing energy   &   108.2             & 0.22           \\
14& L3     &  206.4  &  Missing energy   &   110.1             & 0.21           \\
15& ALEPH  &  206.5  &  Four-jet         &   114.2             & 0.19           \\
16& DELPHI &  206.6  &  Four-jet         &   108.2             & 0.19           \\
17& L3     &  206.6  &  Four-jet         &   109.6             & 0.18           \\
\hline 
\end{tabular}
\caption{\small Properties of the candidates with the largest contribution to $-2\ln Q$ at $\mH =115$~\Gcs. 
The applied selection, $\ln (1+s/b) \ge 0.18$~~(i.e. $s/b \ge 0.2$) at $\mH =115$~\Gcs, retains 17
candidates while the expected numbers of signal 
and background events are 8.4 and 15.8, respectively.
The experiment, the centre-of-mass energy, 
the final state, the reconstructed mass and the weight at $\mH =115$~\Gcs\ are listed. 
\label{tab:event-list}}
\end{center}
\end{table}
For these events,
the evolution of $\ln(1+s/b)$ with test mass is shown in Figure~\ref{fig:spaghetti}. 
Typically, the weight is largest for \mH\ close to the reconstructed mass but there is also a sizeable weight 
over a large domain of test masses due to the experimental resolution, as mentioned before.
For a test mass of 115~\Gcs, the events listed in Table~\ref{tab:event-list} and shown in Figure~\ref{fig:spaghetti} contribute 
with about 40\% to the total weight in the likelihood ratio. 
The distributions of event weights, shown in Figure~\ref{fig:adlo-weights} for two test masses, are in agreement with the 
expectation for the background hypothesis.
\subsection{The reconstructed Higgs boson mass}
\begin{figure}[htb]
\begin{center}
\epsfig{figure=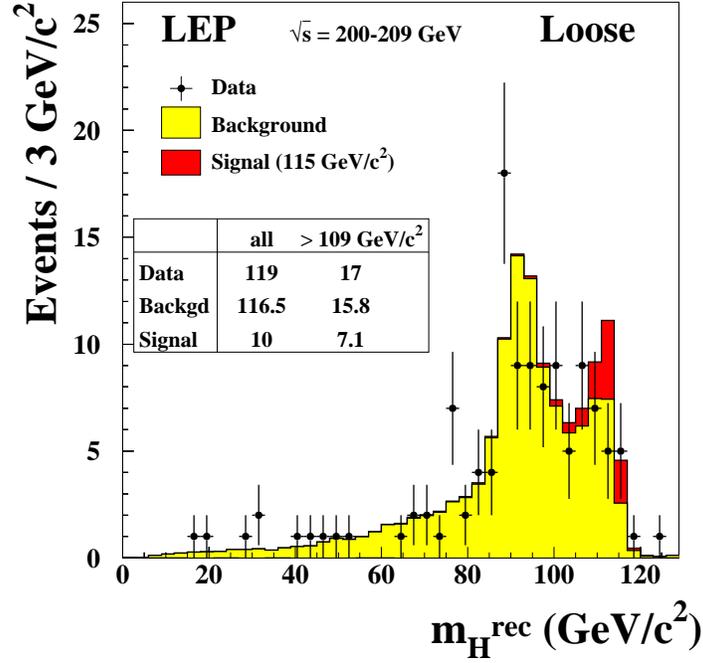,width=0.58\textwidth}\\
\epsfig{figure=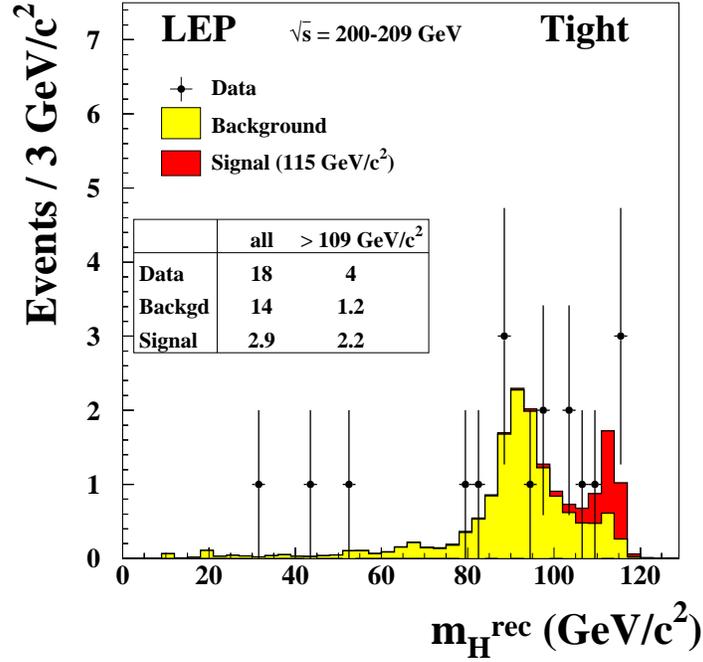,width=0.58\textwidth}\\
\caption[]{\small 
Distributions of the reconstructed Higgs boson mass, $\mHrec$, 
obtained from two selections with differing signal purities. The histograms show the Monte Carlo predictions, lightly shaded for the 
background, heavily shaded for an assumed Standard Model Higgs boson of mass 115~\Gcs, together with the data.
In the loose and tight selections the cuts are adjusted in such a way as to obtain, for a Higgs
boson of mass 115~\Gcs, 
approximately 0.5 or 2 times more expected signal than
background events when integrated over the region $\mHrec > 109$~\Gcs. In the searches where the event selection  
depends on the test mass (see the Appendix), its value is set at 115~\Gcs. 
\label{fig:masses}}
\end{center}
\end{figure}

The reconstructed Higgs boson mass $\mHrec$ is one of the crucial variables which contribute
to the discrimination between the signal and the background and thus to the test statistic $-2\ln Q$.
In Figure~\ref{fig:masses} the distributions for this important
discriminating variable are shown at two levels of signal purity\footnote{These distributions do not enter directly into the 
hypothesis testing but have been produced to illustrate the level of agreement between the data and the 
Monte Carlo simulation.}.  
There is a clear peak in the background prediction in the vicinity of $m_{\rm Z}$ due to the \ee\ra~ZZ background process which is
reproduced by the data. 
\section{Results of the hypothesis testing}
\begin{figure}[htb]
\begin{center}
\epsfig{figure=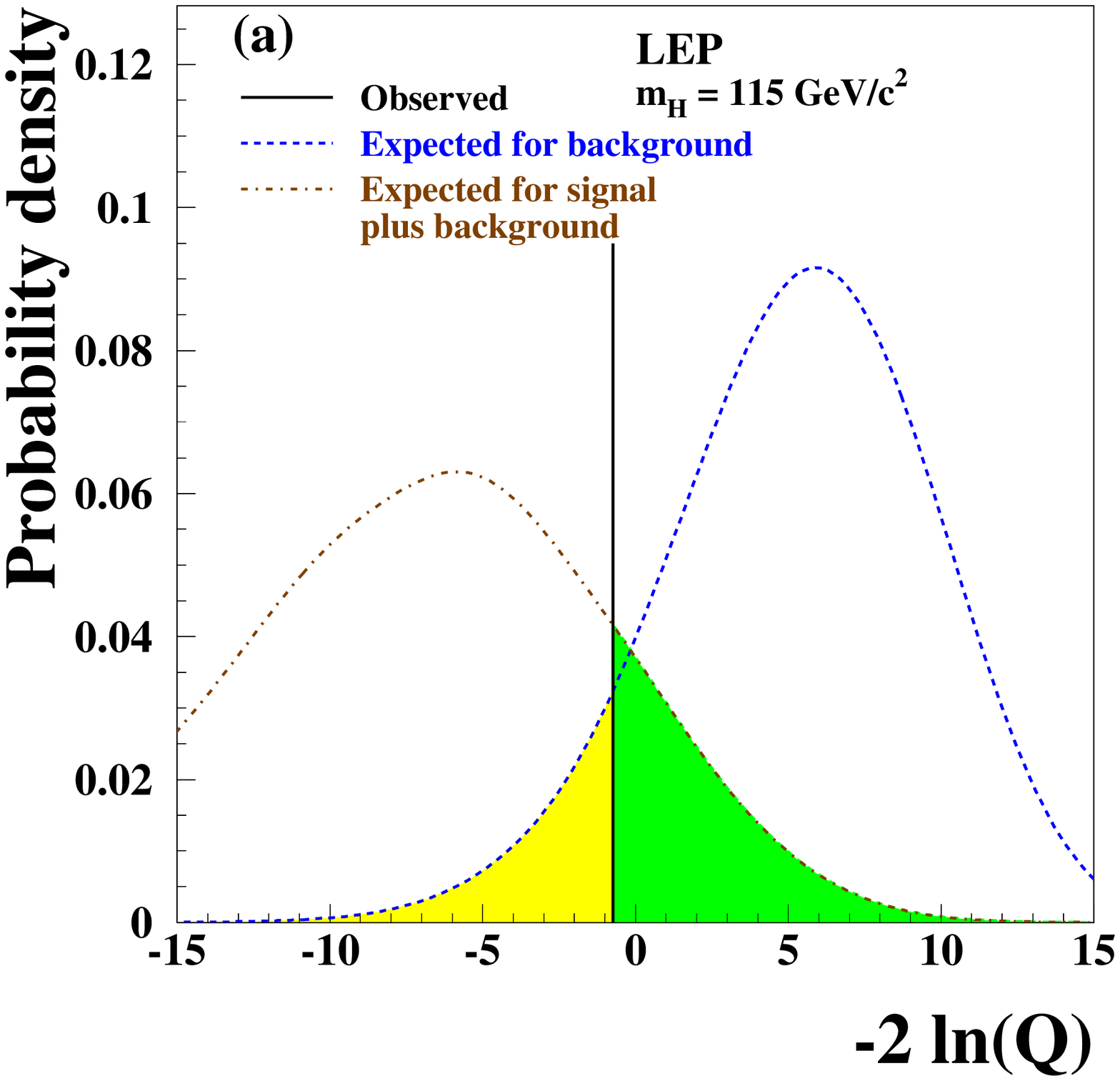,width=0.6\textwidth}\\
\epsfig{figure=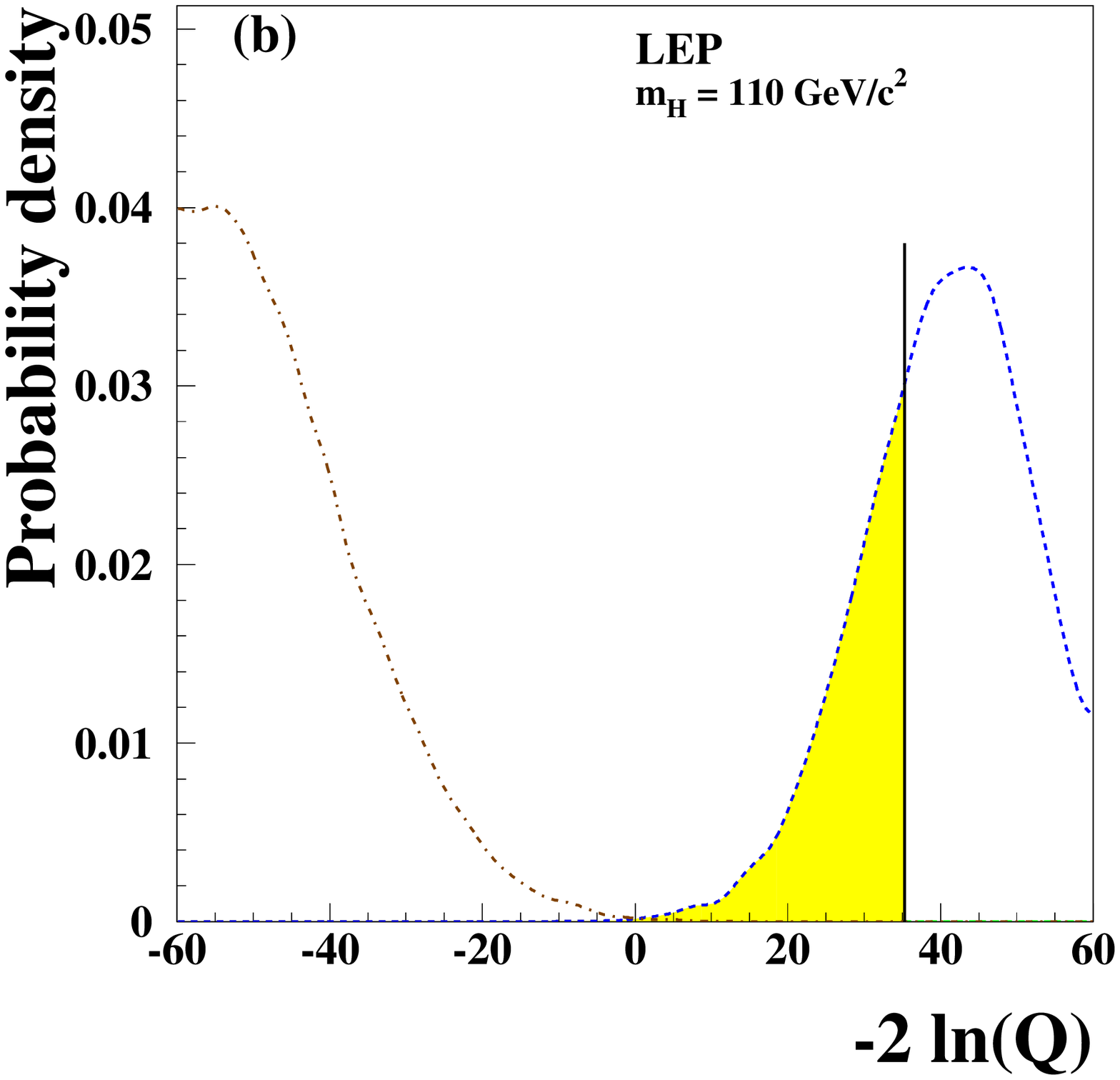,width=0.49\textwidth}
\epsfig{figure=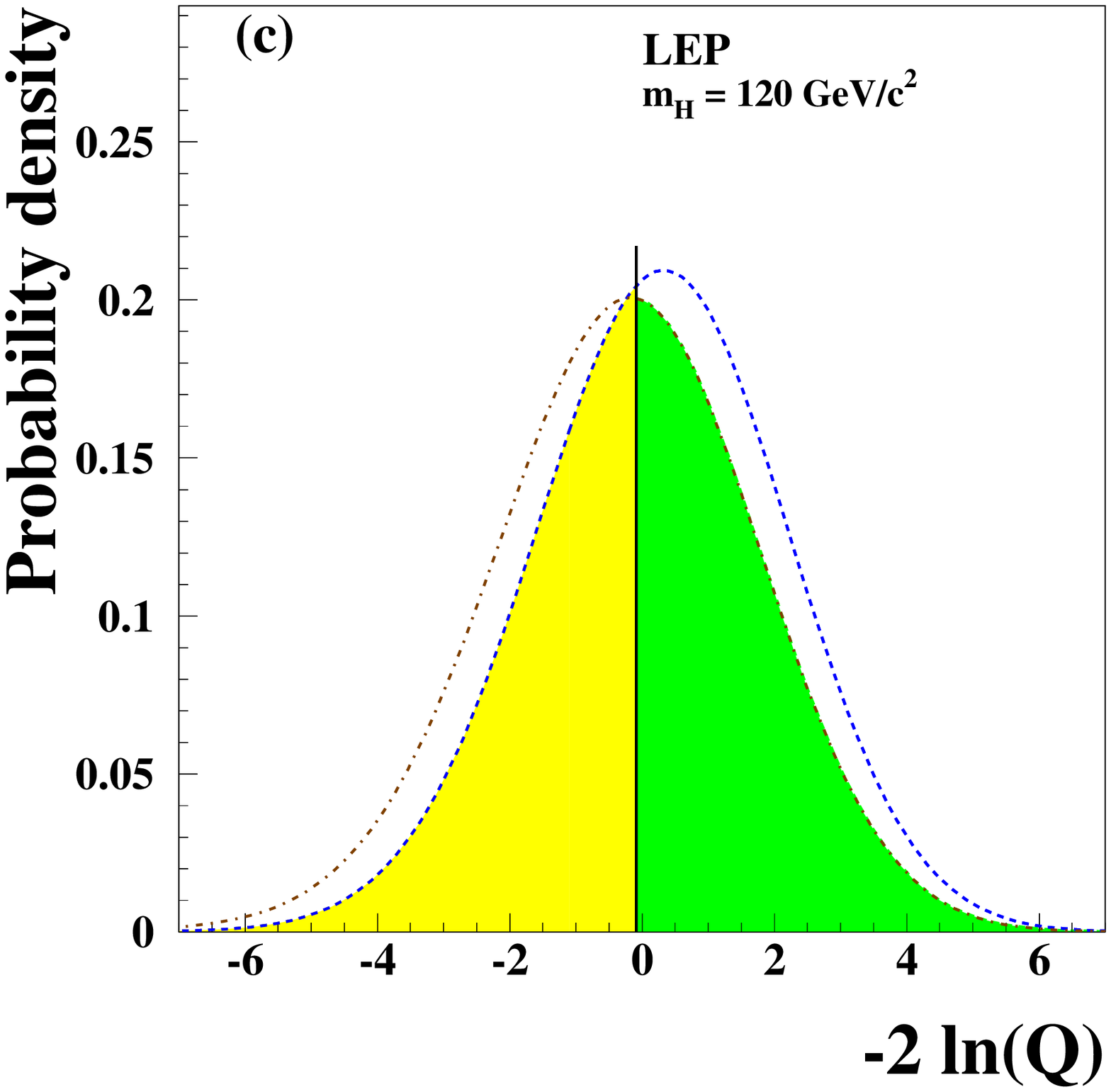,width=0.49\textwidth}
\caption[]{\small Probability density functions corresponding to fixed test masses \mH, for the background 
and signal plus background hypotheses. The observed values of the test statistic $-2\ln Q$ are indicated by the 
vertical lines. The light shaded areas, $1-\clb$, measure the confidence for the background hypothesis 
and the dark shaded areas, $\clsb$, the confidence for the signal plus background hypothesis.
Plot (a): test mass $\mH =115$~\Gcs; (b): $\mH = 110$~\Gcs; (c): $\mH = 120$~\Gcs.
\label{fig:adlo-prob-dens}}
\end{center}
\end{figure}

The expected distributions of the test statistic $-2\ln Q$ from the combined LEP analysis
are shown in Figure~\ref{fig:adlo-prob-dens} for three test masses.
These distributions, which can be thought of as ``slices" of Figure~\ref{fig:adlo-lnq} at the corresponding test masses,
are probability density functions (PDF) for the background and the signal plus background hypotheses and include both the effects
of random statistical variations in the numbers of events
and the systematic uncertainties affecting the expected rates.
Systematic uncertainties are incorporated by varying within the errors
the signal and background estimates in each channel
and evaluating for each variation the probabilities for the data configuration to occur.  
Correlations are addressed by
choosing independent random fluctuations for each source of uncertainty and
applying the changes coherently in those channels
where they apply. 
The uncorrelated errors are dominated by the limited statistics of the simulated background event samples. 
Errors which are correlated between the experiments arise mainly from using the same Monte Carlo generators 
and cross-section calculations, for example for the signal processes. 
The three parts of Figure~\ref{fig:adlo-prob-dens} 
demonstrate a significant discriminating power of the combined LEP data for \mH=110~\Gcs, a moderate one
for \mH=115~\Gcs\ and the rapid decrease of the discriminating power towards the end of the investigated range, \mH=120~\Gcs.

The vertical line in each part of Figures~\ref{fig:adlo-prob-dens} indicates the observed value of $-2\ln Q$ 
for the corresponding test mass.
Integrating the PDF for the background hypothesis from $+\infty$ to the observed value, 
one obtains the background confidence $1-\clb$ which expresses the compatibility of the observation with the background 
hypothesis (also known as p-value, see Ref.~\cite{pdg_stat}). For a large number of simulated measurements 
with no signal and given the background hypothesis,
$1-\clb$ is the probability to obtain a configuration of events which is less background-like (or more signal plus background-like) 
than the one observed.
Similarly, integrating the PDF for the signal plus background hypothesis from the observed value of the test statistic to $-\infty$, 
one obtains the confidence (p-value) $\clsb$ which quantifies the 
compatibility with the signal plus background hypothesis. 

Figure~\ref{fig:adlo-clb} shows the background confidence $1-\clb$ for test masses in the range from 80 to 120~\Gcs. 
In the region \mH$\approx$98~\Gcs\ the value of about 0.02 translates into 2.3  standard deviations (see the Appendix for the conversion).
Note that the number of signal events which would produce such a deviation from the background expectation
is about an order of magnitude smaller than the number expected in the Standard Model for a Higgs boson of this mass.
In the region of \mH\ above 115~\Gcs\ the approximate value of 0.09  
translates into 1.7 standard deviations from the background hypothesis.
This deviation, although of low significance, is compatible with
a Standard Model Higgs boson in this mass range while being also in agreement with the background hypothesis. 
Note that the value of $1-\clb$ would change in this region from about 0.09 to about 0.08 if the systematic errors were ignored.

The dash-dotted line in Figure~\ref{fig:adlo-clb} shows the position of the minimum of the median expected $1-\clb$ 
for the signal plus background hypothesis, when the signal mass given on the abscissa is tested. This line is an indication for 
the limiting sensitivity of the combined LEP data for detecting a Higgs boson signal.
For example, in the case of a Higgs boson of mass 115~\Gcs, a minimum of about 0.008 would be expected. 

Figure~\ref{fig:clb-subsets}~shows $1-\clb$ as a function of the test mass for subsets of the LEP data.
The confidences $1-\clb$ and $\clsb$, for a test mass of 115~\Gcs, are
listed in Table~\ref{tab:clb} for all LEP data combined and for various sub-samples.  
\begin{table}[t]
\begin{center}
\begin{tabular}{|l|cc|}
\hline
                  &      $1-\clb$             &   $\clsb$        \\
\hline\hline
LEP               &         0.09              &    0.15         \\
\hline
ALEPH             &   $3.3\times 10^{-3}$     &    0.87         \\
DELPHI            &         0.79              &    0.03          \\
L3                &         0.33              &    0.30          \\
OPAL              &         0.50              &    0.14          \\
\hline
Four-jet          &         0.05              &    0.44        \\
All but four-jet  &         0.37              &    0.10        \\ 
\hline
\end{tabular}
\caption{\small The background confidence $1-\clb$  and the signal plus background 
confidence $\clsb$ for a test mass $\mH =115$~\Gcs, for all LEP data combined and for various 
subsets. The values for the four-jet and all but the four-jet final states are obtained 
with the data of the four experiments combined.
\label{tab:clb}}
\end{center}
\end{table}
\section{Bounds for the Higgs boson mass and coupling}
The ratio $\cls = \clsb /\clb$ as a function of the test mass, shown in 
Figure~\ref{fig:adlo-cls}, is used to derive a lower bound on the Standard Model Higgs boson mass (see the Appendix). 
The lowest test mass giving $\cls=0.05$ is taken as the lower bound on the mass at the 95\% confidence level.
\begin{table}[t]
\begin{center}
\begin{tabular}{|l|cc|}
\hline
                               & Expected limit (\Gcs) & Observed limit (\Gcs)    \\
\hline\hline
LEP                            &    115.3             &   114.4   \\
\hline
ALEPH                          &    113.5             &   111.5   \\   
DELPHI                         &    113.3             &   114.3   \\ 
L3                             &    112.4             &   112.0   \\ 
OPAL                           &    112.7             &   112.8    \\
\hline
Four-jet channel               &    114.5             &   113.3   \\ 
All but four-jet               &    114.2             &   114.2   \\ 
\hline
\end{tabular}
\caption{\small Expected (median)  and observed 95\% confidence level lower bounds on the Standard Model Higgs boson mass, 
for all LEP data combined and for various 
subsets of the data.
The numbers for the four-jet and all but the four-jet final states are obtained with the data of the four experiments combined.
\label{tab:mass-limits}}
\end{center}
\end{table}
The expected and observed lower bounds are listed in Table~\ref{tab:mass-limits}. The expected limits provide an indication of 
the relative sensitivities of the data subsets.
The observed, combined, 95\% confidence level lower bound on the mass of the Standard Model Higgs boson obtained at LEP is 114.4~\Gcs,
while the median expected limit is 115.3 \Gcs. The difference reflects the slight excess of the data with respect to the background expectation
at high masses. The observed and the expected limits would shift upwards by about 50~MeV/$c^2$ if the systematic errors were ignored. 

\begin{figure}[htb]
\begin{center}
\epsfig{figure=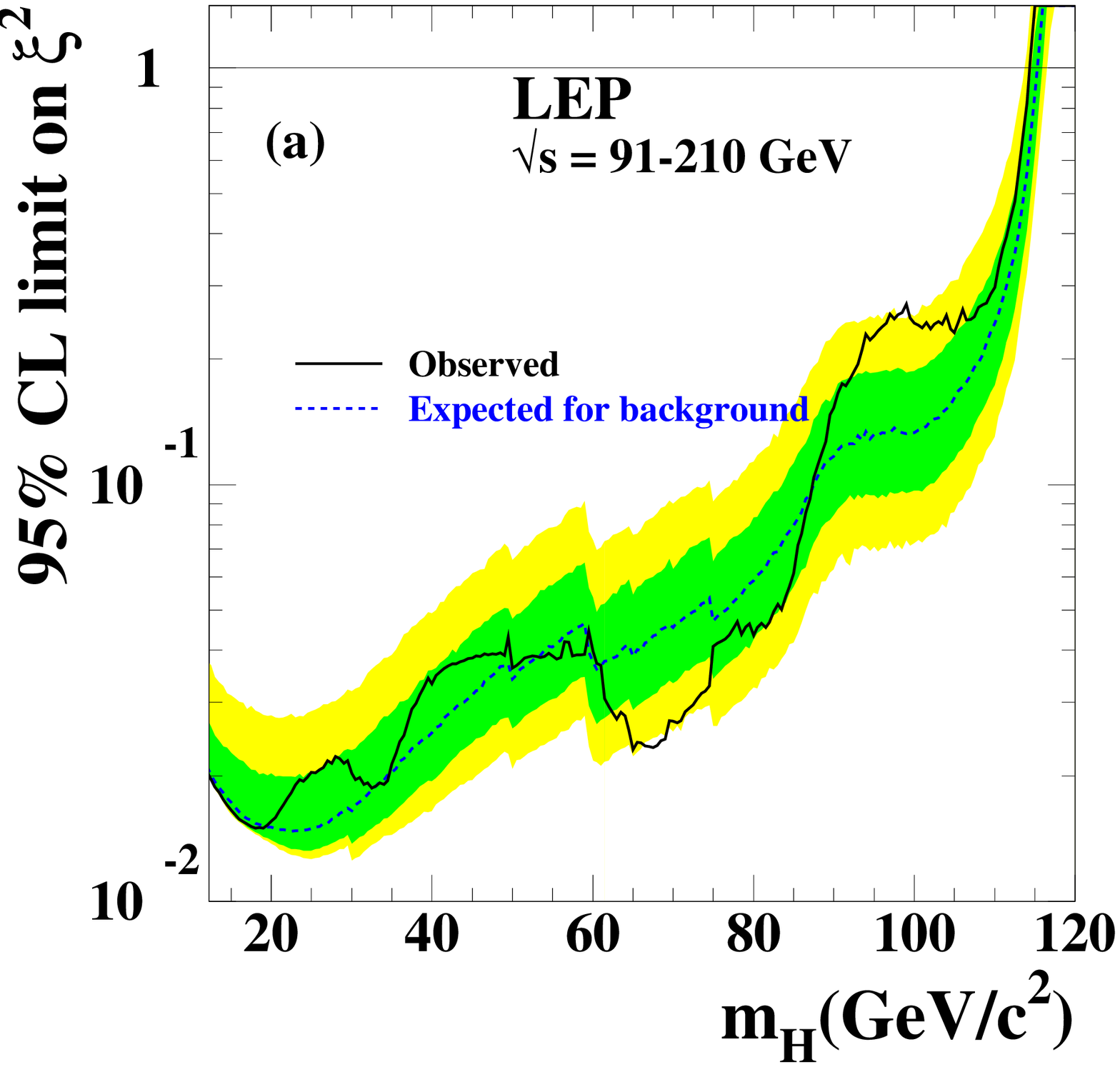,width=0.49\textwidth} \\
\epsfig{figure=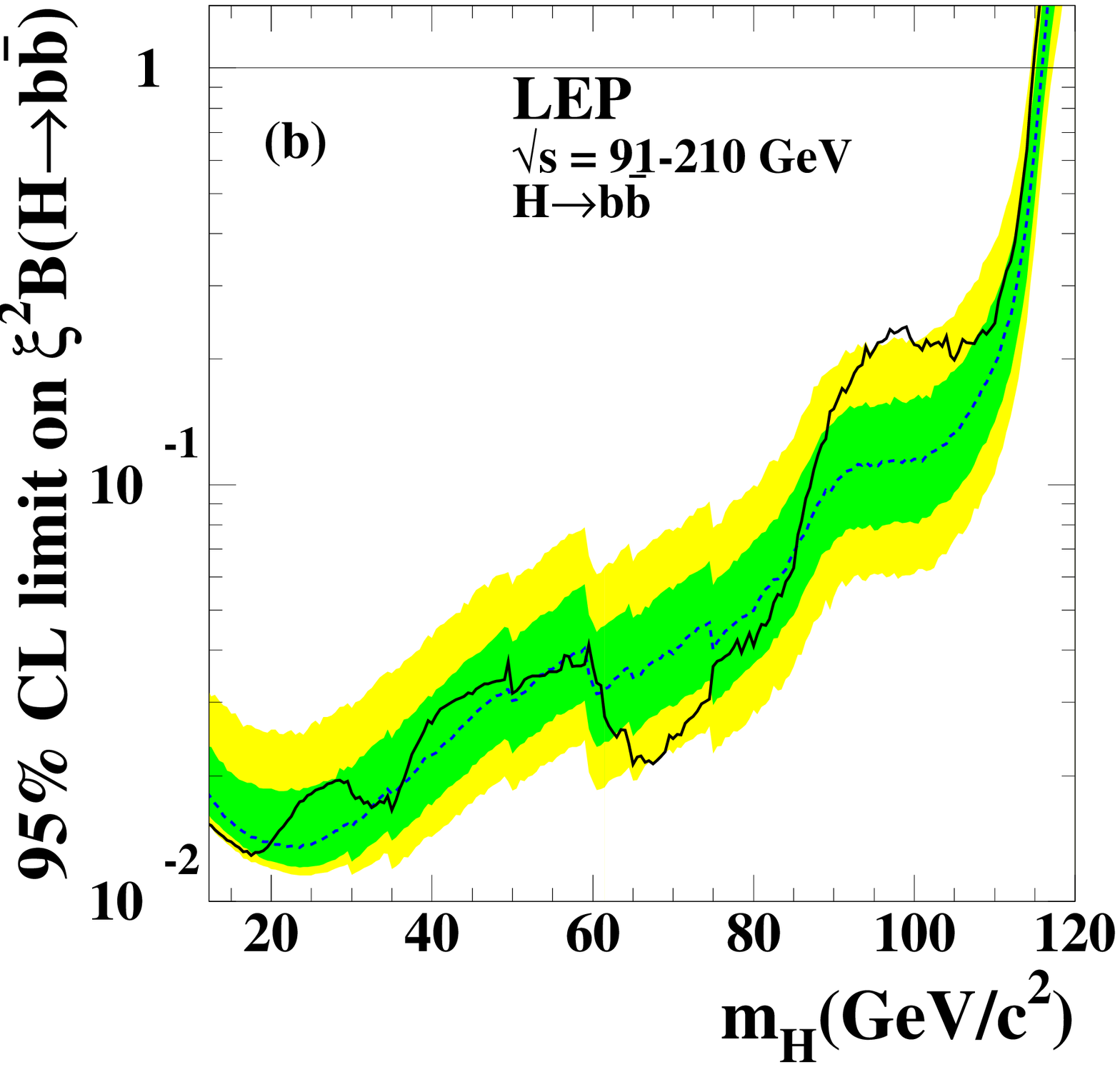,width=0.49\textwidth} 
\epsfig{figure=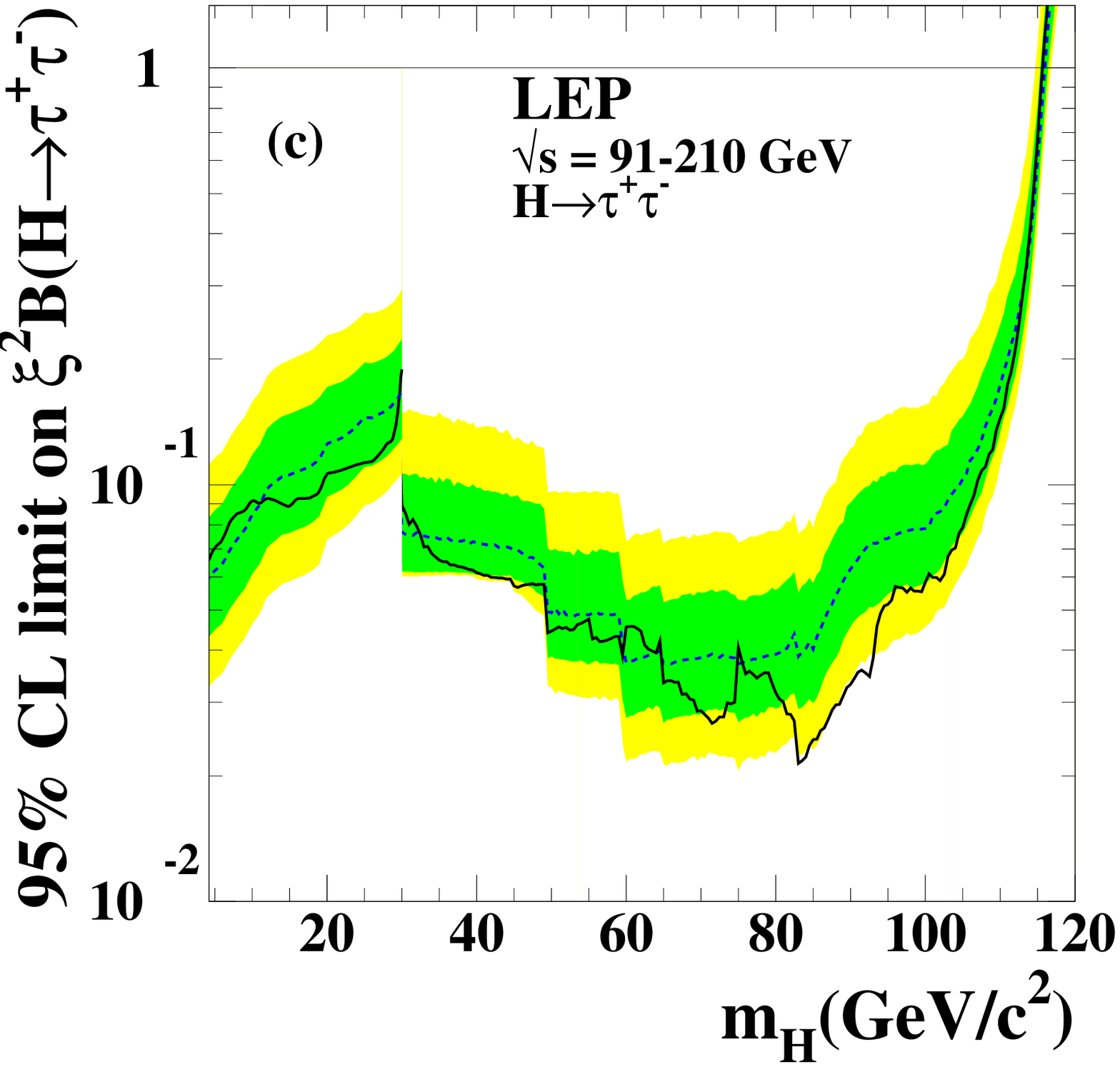,width=0.49\textwidth}\\ 
\caption[]{\small The 95\% confidence level upper bound on the ratio $\xi^2 = (g_{\rm HZZ}/g_{\rm HZZ}^{\rm SM})^2$ (see text). 
The dark and light shaded bands around the median
expected line correspond to the 68\% and 95\% probability bands.
The horizontal lines correspond to the Standard Model coupling. 
(a): For Higgs boson decays predicted by the Standard Model; (b): 
for the Higgs boson decaying exclusively into \bb\ and (c): into $\tau^+\tau^-$ pairs. 
\label{sm-xi2}}
\end{center}
\end{figure}

The combined LEP data are also used to set 95\% confidence level upper bounds on the HZZ coupling in non-standard models.
In the ratio \mbox{$\xi^2= (g_{\rm HZZ}/g_{\rm HZZ}^{\rm SM})^2$} the variable $g_{\rm HZZ}$ designates the non-standard HZZ coupling 
and $g_{\rm HZZ}^{\rm SM}$ the same coupling in the Standard Model.
In deriving the limits on $\xi^2$, which cover a broad range of masses, the LEP1 data collected at the Z resonance~\cite{lep1}
have been combined with LEP2 data taken at energies between 161 and 209~GeV. In part (a) of Figure~\ref{sm-xi2} 
the Higgs boson is assumed to decay into fermions and bosons according to the Standard Model while the 
cross-sections for the process \ee\ra~HZ and the fusion processes WW\ra~H and ZZ\ra~H are scaled with $g_{\rm HZZ}^2$.
For masses below 12~\Gcs, not shown in the figure, the limits quoted in~\cite{low-a,low-d,low-l,low-o} apply.
In parts (b) and (c) it is assumed that the Higgs boson decays exclusively into \bb\ or $\tau^+\tau^-$ pairs.
In the $\tau^+\tau^-$ case and for masses below 30~\Gcs, the limit shown is provided by the search of Ref.~\cite{decindep-o}.
\section{Summary}
Combining the final results from the four LEP experiments, ALEPH, DELPHI, L3 and OPAL,
a lower bound of 114.4~\Gcs\ is set on the mass of the Standard Model Higgs boson at the 95\% confidence
level. At the beginning of the LEP programme no solid limit existed for the mass of this particle~\cite{franzini}.

At a mass of 115~\Gcs, where ALEPH reported an excess 
compatible with the production of the Standard Model Higgs boson, the 
confidence $1-\clb$ of the combined LEP data expressing the level of consistency with the 
background hypothesis is 0.09, while the confidence $\clsb$ measuring the consistency with the signal plus 
background hypothesis is 0.15. 

The LEP1 and LEP2 data have been used in conjunction to set upper bounds
on the HZZ coupling for a wide range of Higgs boson masses and for various assumptions concerning the Higgs boson
decay properties.
 
The searches for the Standard Model Higgs boson carried out by the four LEP experiments 
extended the sensitive range well beyond that anticipated at the beginning of the 
LEP programme~\cite{lep-86}. This is due to the higher energy achieved and to more sophisticated detectors and
analysis techniques.

\newpage
\vspace{2cm}
\noindent{\Large\bf Acknowledgements}\\

\noindent 
We congratulate our colleagues from the CERN Accelerator Divisions for the successful running of the LEP collider, particularly in the year 2000 at 
the highest energies. We would also like to express our thanks
to the engineers and technicians in all our institutions for their contributions to the excellent
performance of the four LEP detectors. 
The LEP Working Group for Higgs Boson Searches acknowledges the fruitful cooperation between the
collaborations in developing the combination procedures and applying them to the LEP data.

\vspace{1cm}
\noindent{\Large\bf Appendix:~~Statistical method}

The test statistic adopted in the combination of LEP data~\cite{junk} is $-2\ln Q$ where $Q$ is the ratio of the likelihood function 
for the signal plus background hypothesis to the likelihood function for the background hypothesis,
   \begin{equation}
       Q(m_{\rm H})= \frac{{\cal L}_{\rm s+b}}{{\cal L}_{\rm b}}.
   \end{equation}
The binned likelihood functions are defined by
   \begin{equation}
      {\cal L}(\eta)=
      \prod_{k=1}^{N} {\exp[-(\eta s_k(m_{\rm H})+b_k)]
      ~(\eta s_k(m_{\rm H})+b_k)^{n_k} \over n_k!} \times            
      \prod_{j=1}^{n_k}{\eta s_k(m_{\rm H})S_k(\vec{x}_{jk}; m_{\rm H})
      +b_kB_k(\vec{x}_{jk})\over \eta s_k(m_{\rm H})+b_k},
   \end{equation}
where $\eta = 1$ in the case of ${\cal L}_{\rm s+b}$ and $\eta = 0 $ in the case of ${\cal L}_{\rm b}$.
The index $k$ runs over all independent contributions to the combined result: from different event topology selections, data
taken at different centre-of-mass energies and data collected in different experiments.
The symbol $N$ stands for the number of such contributions (``channels");
$n_k$ is the number of 
observed candidates in channel $k$ and $\vec{x}_{jk}$ designates the position $\vec{x}$ 
of candidate $j$ of channel $k$ in the plane defined by the discriminating variables $\mHrec$ and $\cal G$ (see Section
2). The quantities
$s_k(m_{\rm H})$ and $b_k$ are the integrated signal and background rates in
channel $k$.
The functions
$S_k(\vec{x}_{jk}; m_{\rm H})$ and $B_k(\vec{x}_{jk})$ are the probability density functions (PDFs) of the discriminating 
variables
for the signal and background. These PDFs are evaluated in bins of \mHrec\ and $\cal G$ for a set
of values for \mH\ with some interpolation and smoothing procedures
applied~\cite{interpol,keys}. The test statistic can be written
\begin{equation}
-2\ln Q (m_{\mathrm H}) = 2\sum_{k=1}^N\left[s_k(m_{\mathrm{H}}) -
 \sum_{j=1}^{n_k}\ln \left(1+
\frac{s_k(m_{\mathrm{H}})S_k(\vec{x}_{jk};m_{\mathrm{H}})}
     {b_kB_k(\vec{x}_{jk})}
 \right)\right],
\end{equation}
thus becoming a sum of contributions (weights) from the
individual observed events.The above notation assumes that
the background-related quantities $b_k$ and $B_k(\vec{x}_{jk})$ do not depend on \mH. 
If the selection criteria in any one channel are explicitly \mH\
dependent (the searches of L3 and OPAL in the four-jet channel have this property), 
$b_k$ and $B_k(\vec{x}_{jk})$ have to be replaced by $b_k(m_{\rm H})$ 
and $B_k(\vec{x}_{jk}; m_{\rm H})$.

The presence of a signal can be inferred from the behaviour of the confidence $1-\clb$
for the background hypothesis (also called p-value, see Ref.~\cite{pdg_stat}), which is obtained,
for a given test mass, by integrating the corresponding PDF from 
$-\infty$ to the observed value of the test statistic. The PDFs are obtained from detailed simulations of experiments, 
given the background hypothesis.
If the background hypothesis is correct, $1-\clb$ is uniformly distributed between zero and one; the median of the distribution 
would thus be 0.5. In the presence of a significant signal $1-\clb$ would be very small for the corresponding test mass.

\begin{figure}[htb]
\begin{center}
\epsfig{figure=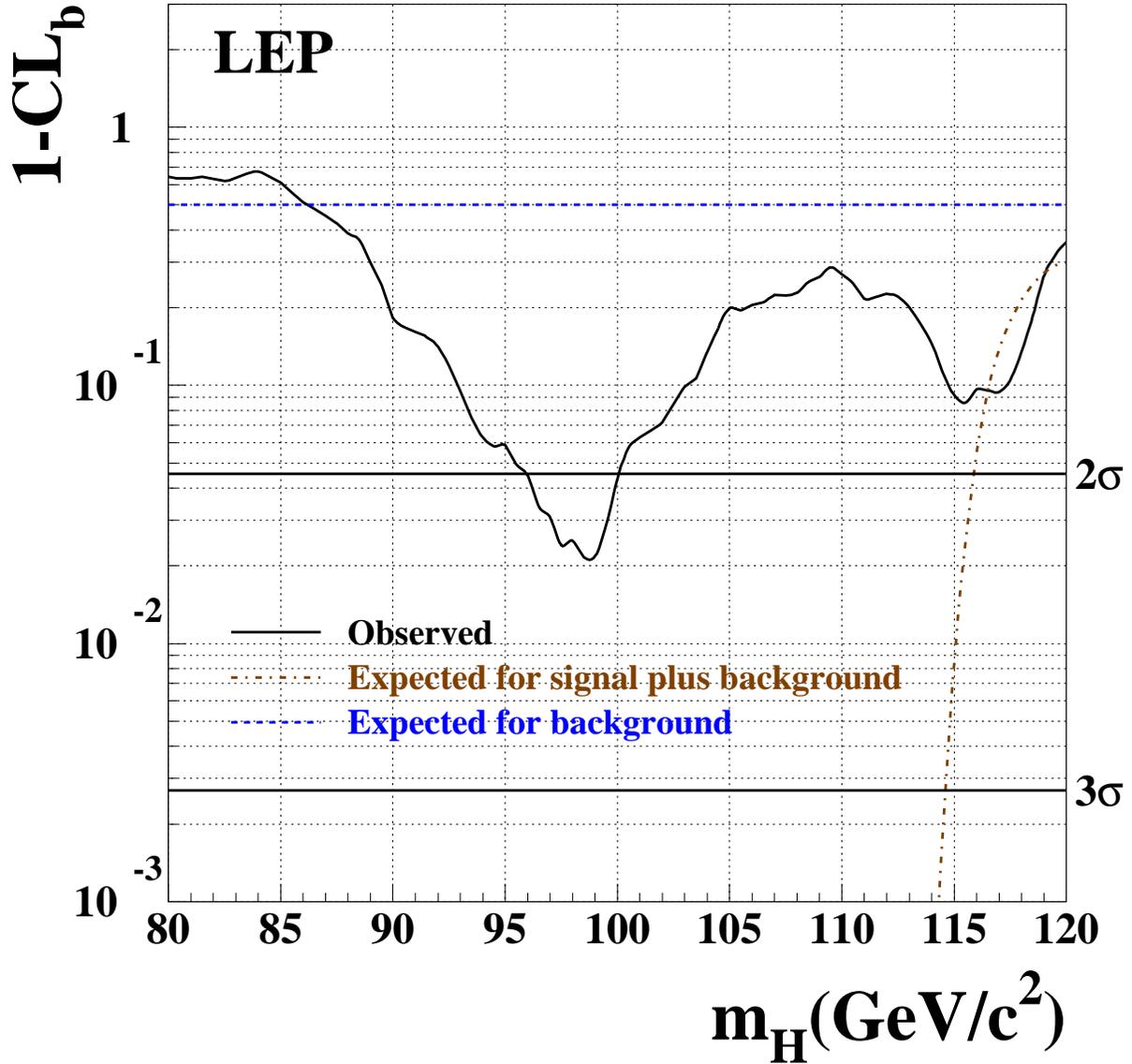,width=\textwidth}\\
\caption[]{\small The background confidence $1-\clb$ as a function of the test mass $\mH $. Full curve: observation; 
dashed curve: expected background confidence; 
dash-dotted line: the position of the minimum of the median expectation of $1-\clb$ for the signal plus background 
hypothesis, when the signal
mass indicated on the abscissa is tested.
The horizontal solid lines indicate the levels for
2$\sigma$ and 3$\sigma$ deviations from the background hypothesis (see the Appendix for the conversion). 
\label{fig:adlo-clb}}
\end{center}
\end{figure}

\begin{figure}[htb]
\begin{center}
\epsfig{figure=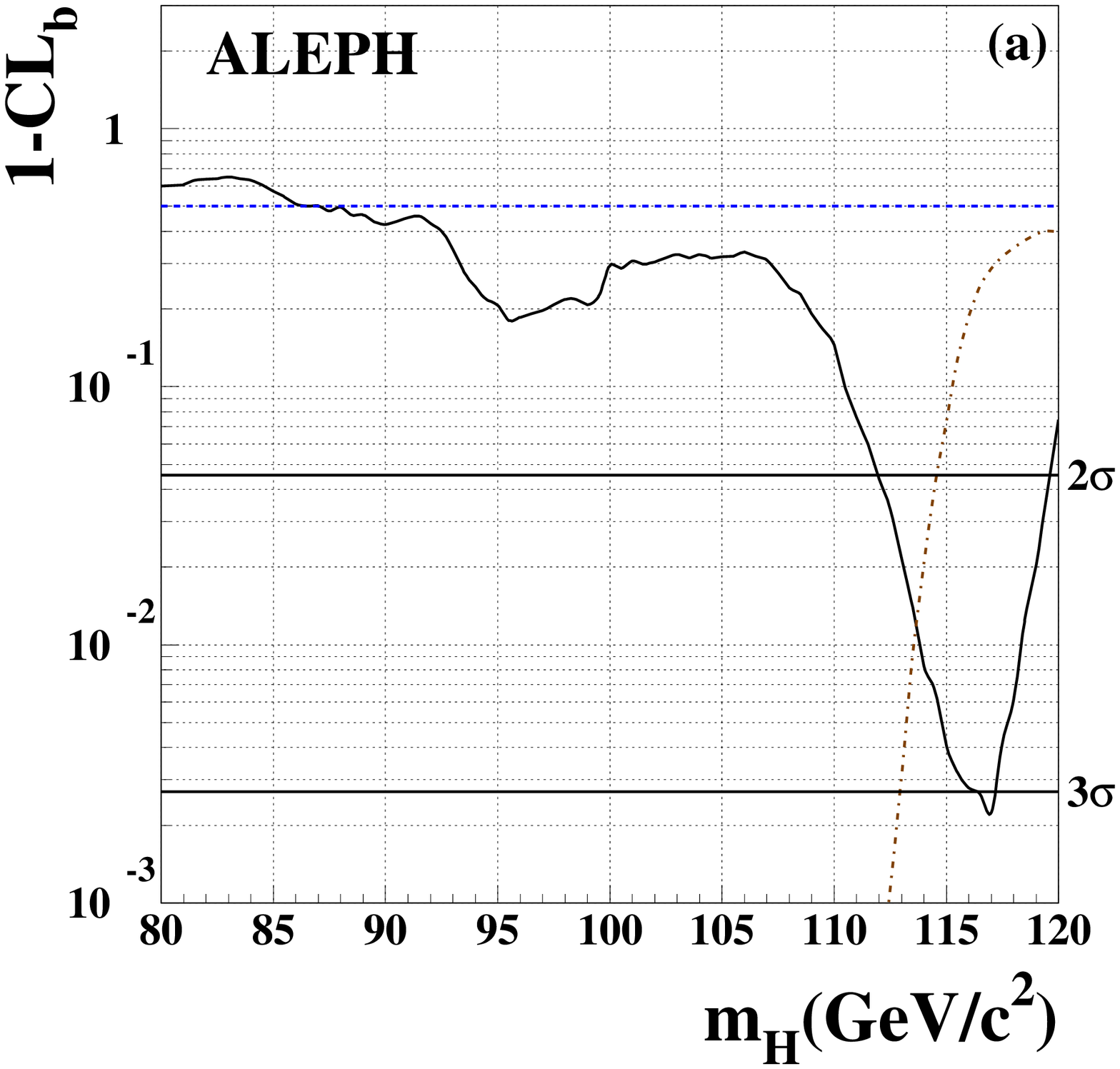,width=0.40\textwidth}
\epsfig{figure=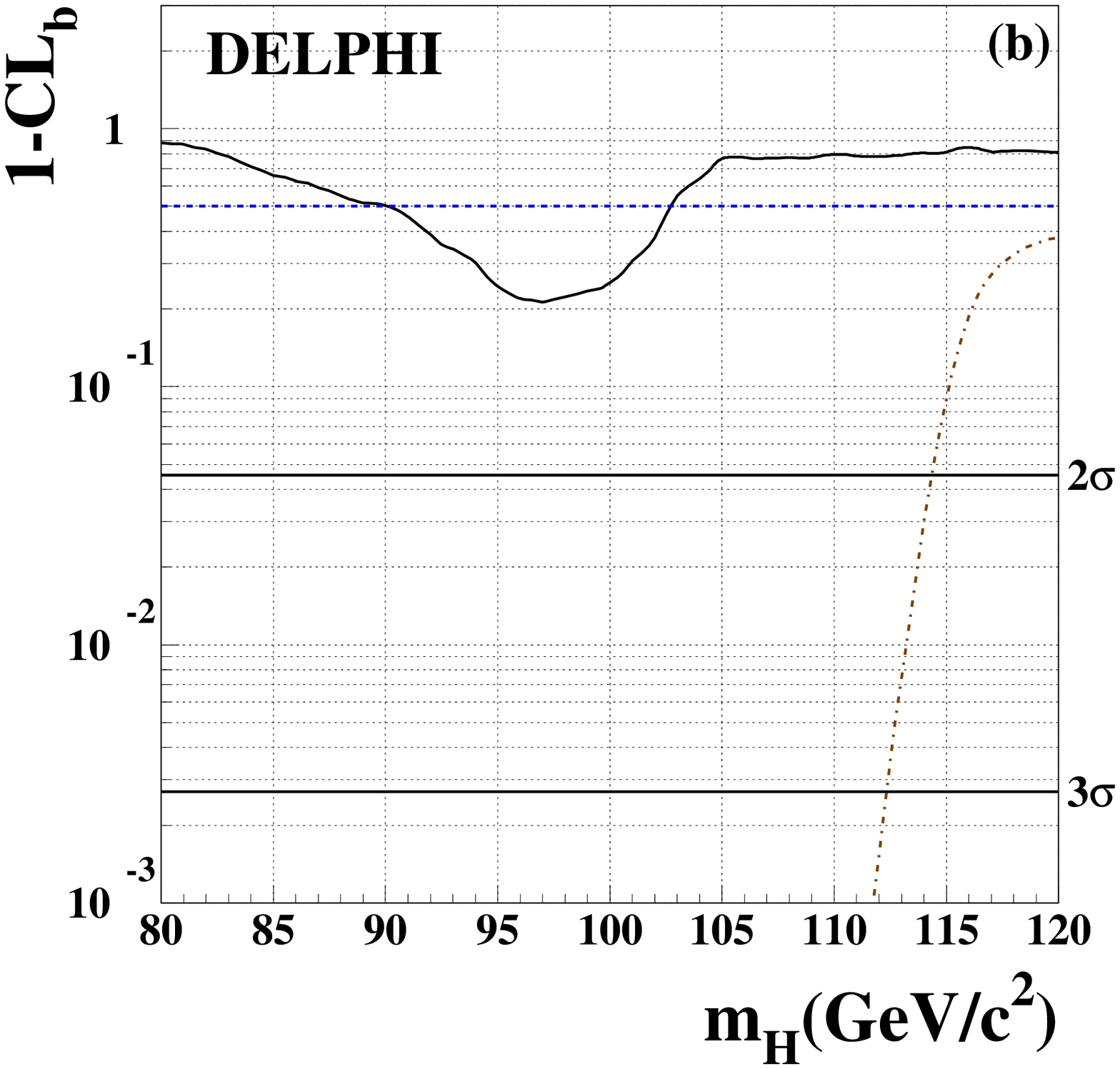,width=0.40\textwidth}\\
\epsfig{figure=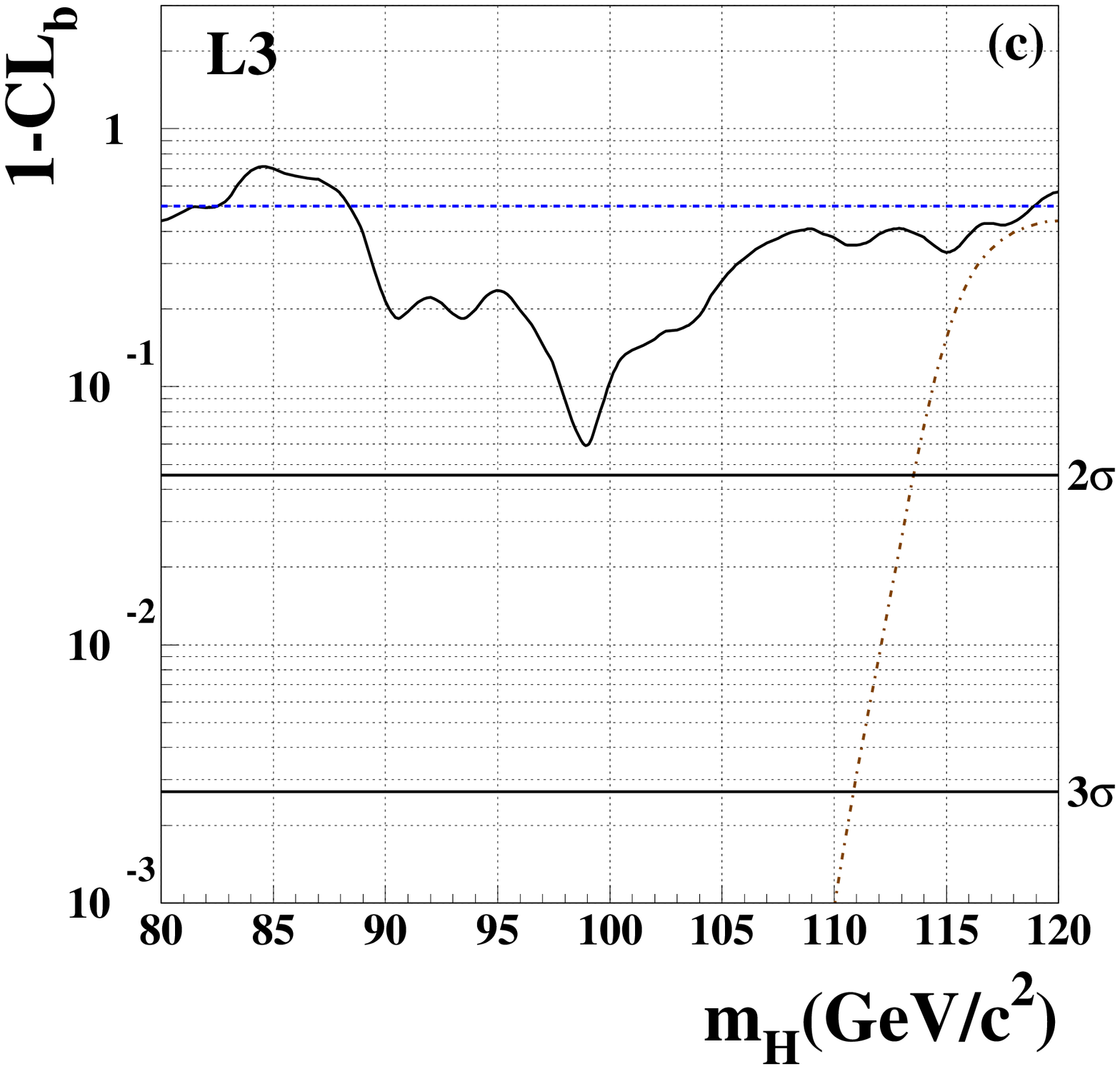,width=0.40\textwidth}
\epsfig{figure=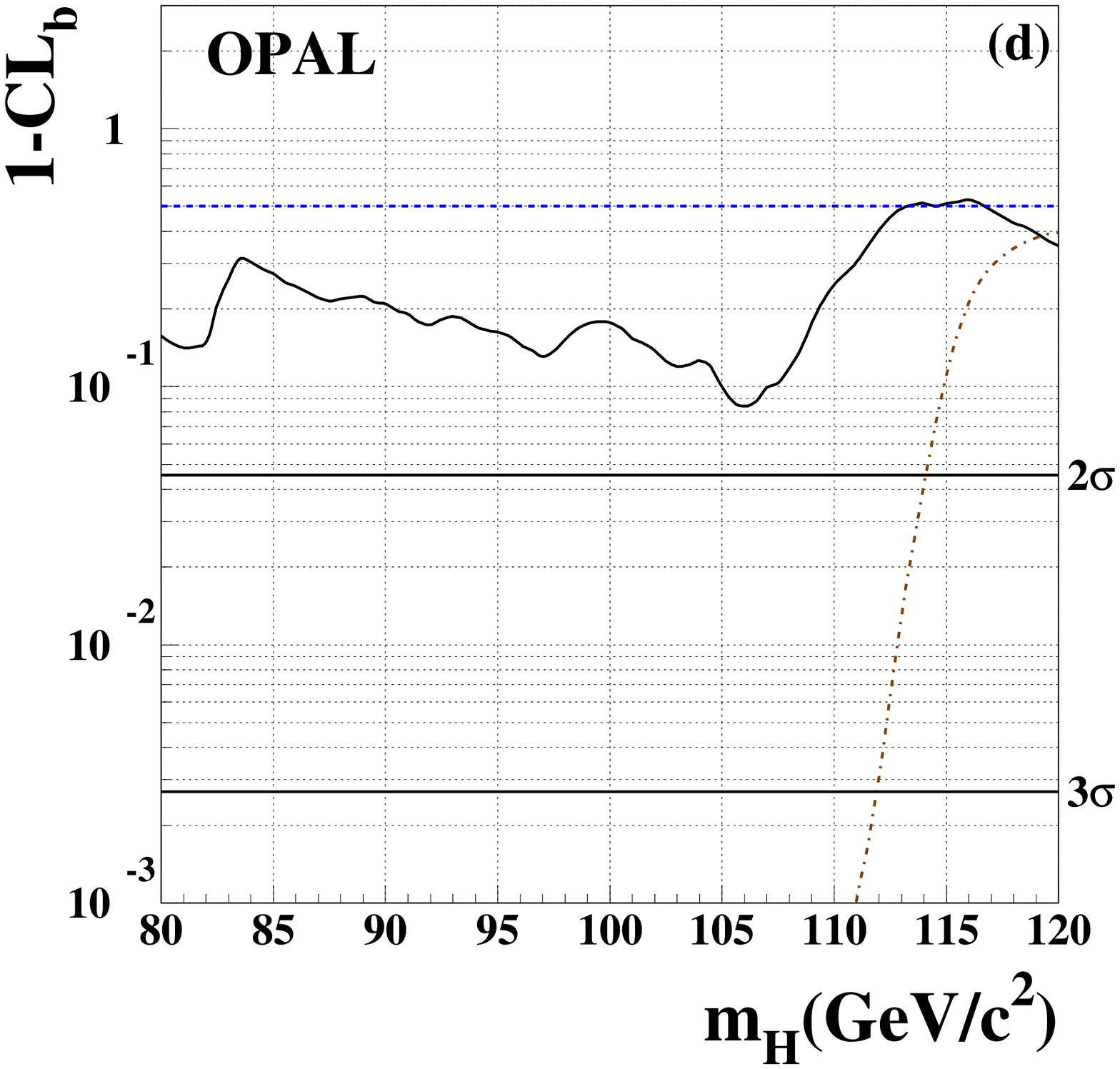,width=0.40\textwidth}\\
\epsfig{figure=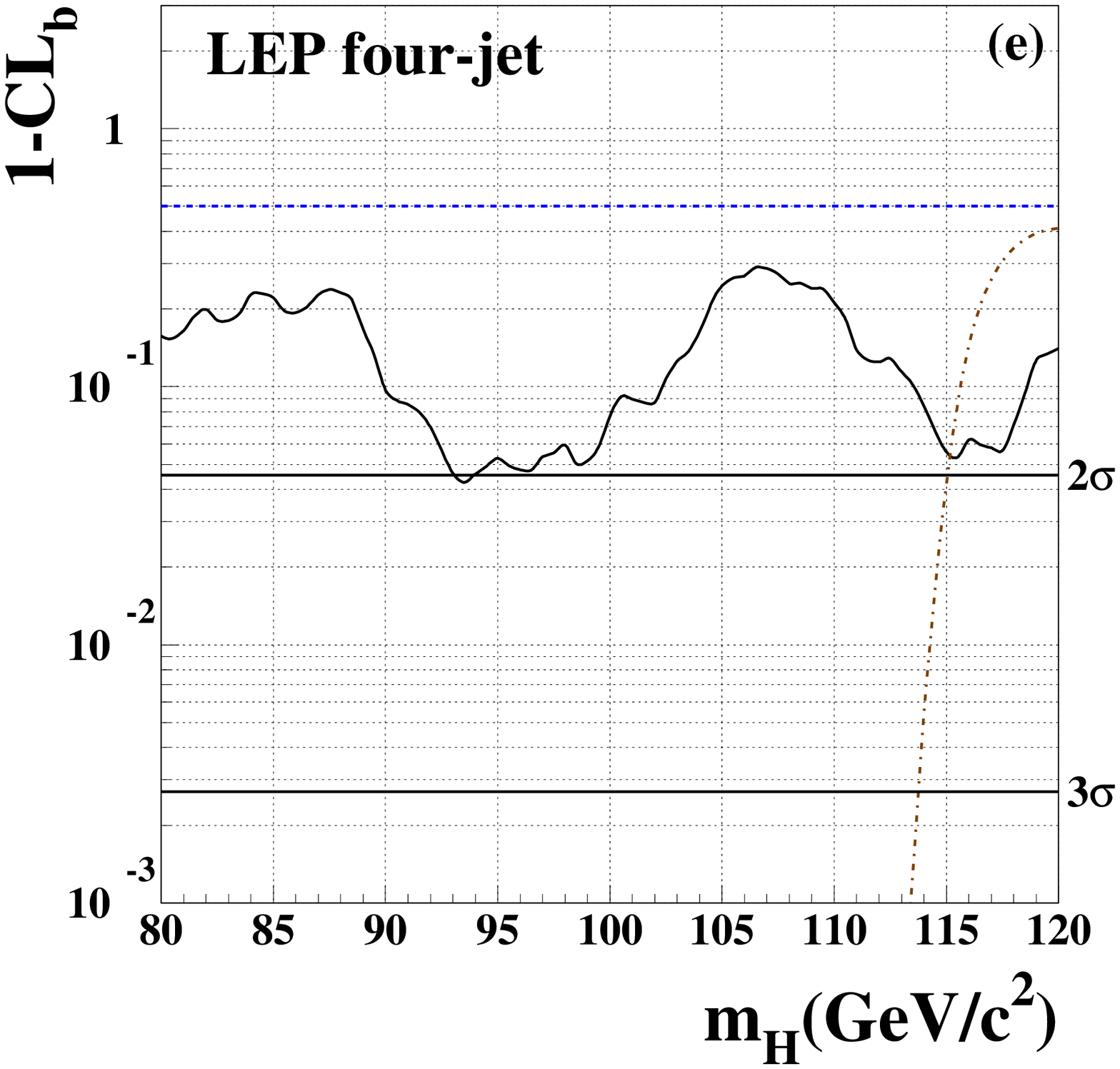,width=0.40\textwidth}
\epsfig{figure=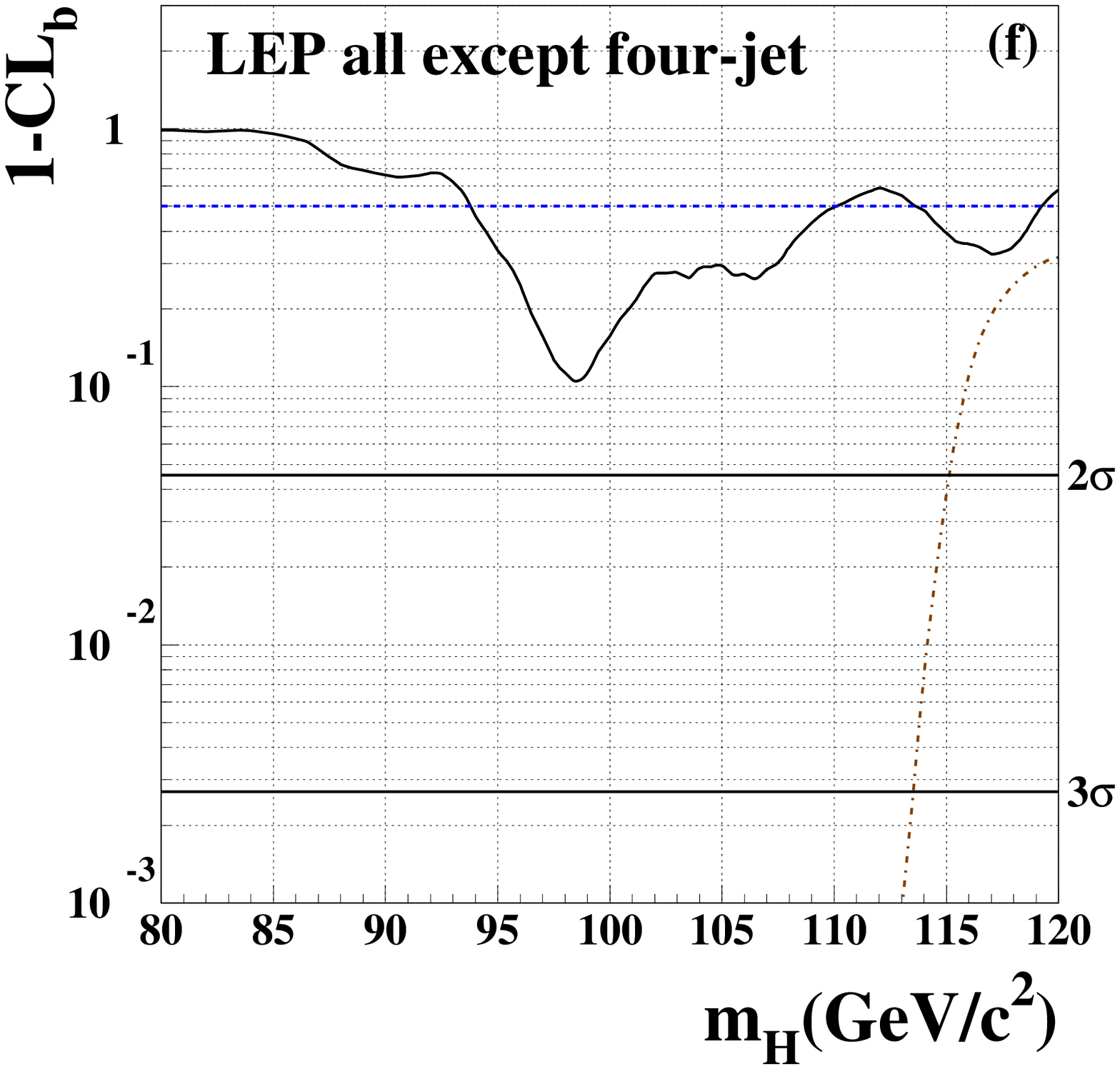,width=0.40\textwidth}\\
\caption[]{\small The background confidence $1-\clb$ as a function of the test mass $\mH $ for subsets of the LEP data.
The same notation as in Figure~\ref{fig:adlo-clb} is used.
Plots (a) to (d): individual experiments; (e): the four-jet and (f): all but the four-jet final state, 
with the data of the four experiments combined.
\label{fig:clb-subsets}}
\end{center}
\end{figure}

To express a given value of $1-\clb$  in terms of standard deviations ($\sigma$), 
we adopt a convention (see Table 31.1 of Ref.~\cite{pdg_stat}) where $1-\clb = 2.7\times 10^{-3}$ ($5.7\times 10^{-7}$) would 
indicate a $3\sigma$ ($5\sigma$) excess beyond the background expectation. The vertical scales on the right-hand side in
Figures~\ref{fig:adlo-clb} and~\ref{fig:clb-subsets} correspond to this convention.

The frequentist exclusion limit is usually computed from the confidence \clsb\ for the signal plus background hypothesis
which, for a given test mass, is obtained by integrating the corresponding PDF  
from the observed value of the test statistic to $+\infty$. The signal plus background hypothesis is considered excluded 
at the 95\% confidence level if 
an observation is made such that \clsb\ is lower than 0.05. However, this procedure may lead to the undesired possibility that a large downward
fluctuation of the background would allow hypotheses to be excluded for which the experiment has no sensitivity due 
to the small expected signal rate. 
This problem is avoided by introducing the ratio \cls\ = \clsb/\clb. Since \clb\ is a positive number less than one, 
\cls\ will always be greater than \clsb\ and the limit obtained will therefore be conservative. 
We adopt this quantity for setting exclusion limits and consider a hypothesis to be excluded 
at the 95\% confidence level if the corresponding value of \cls\ is less than~0.05. 

%%%%%%%%%%%%%%%%%%%%%%%%%%%%%%%%%%%%%%%%%%%%%%%%%%%%%%%%%%%%%%%%%%%%%%%%%%%%%

%**********************************************************************************************
\newpage
\begin{figure}[htb]
\begin{center}
\epsfig{figure=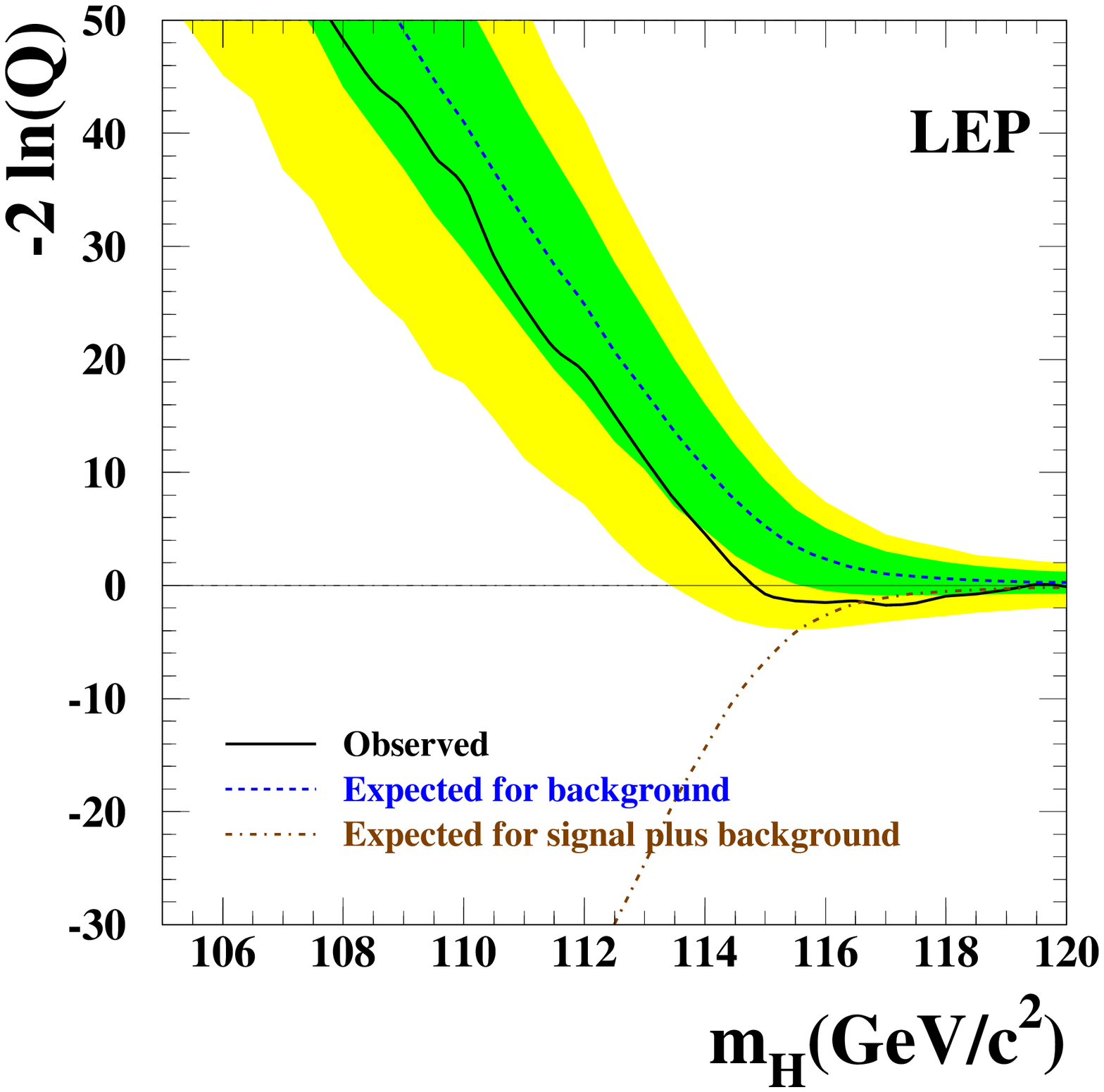,width=0.9\textwidth}\\
\caption[]{\small  Observed and expected behaviour of the test statistic $-2\ln Q$ as a function of the test mass
$\mH $, obtained by combining the data of the four LEP experiments. 
%Upper part: 
The full curve represents the observation; the dashed curve shows the median 
background expectation; the dark and light shaded bands represent the 
68\% and 95\% probability bands about the median background expectation.
The dash-dotted curve indicates the position of the minimum of the median expectation for the signal plus background hypothesis when the signal
mass given on the abscissa is tested.
\label{fig:adlo-lnq}}
\end{center}
\end{figure}
\begin{figure}[htb]
\begin{center}
\epsfig{figure=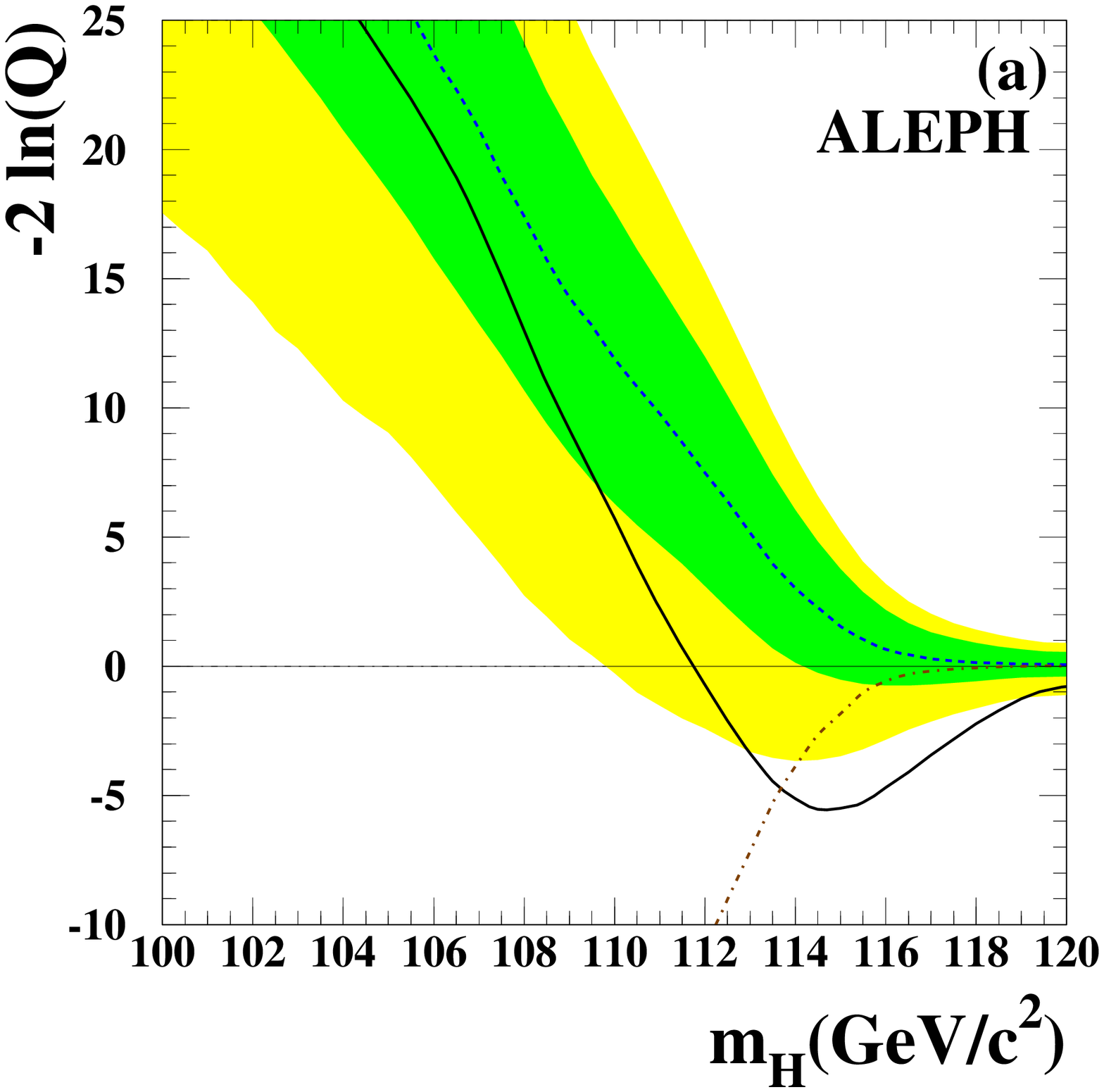,width=0.40\textwidth}
\epsfig{figure=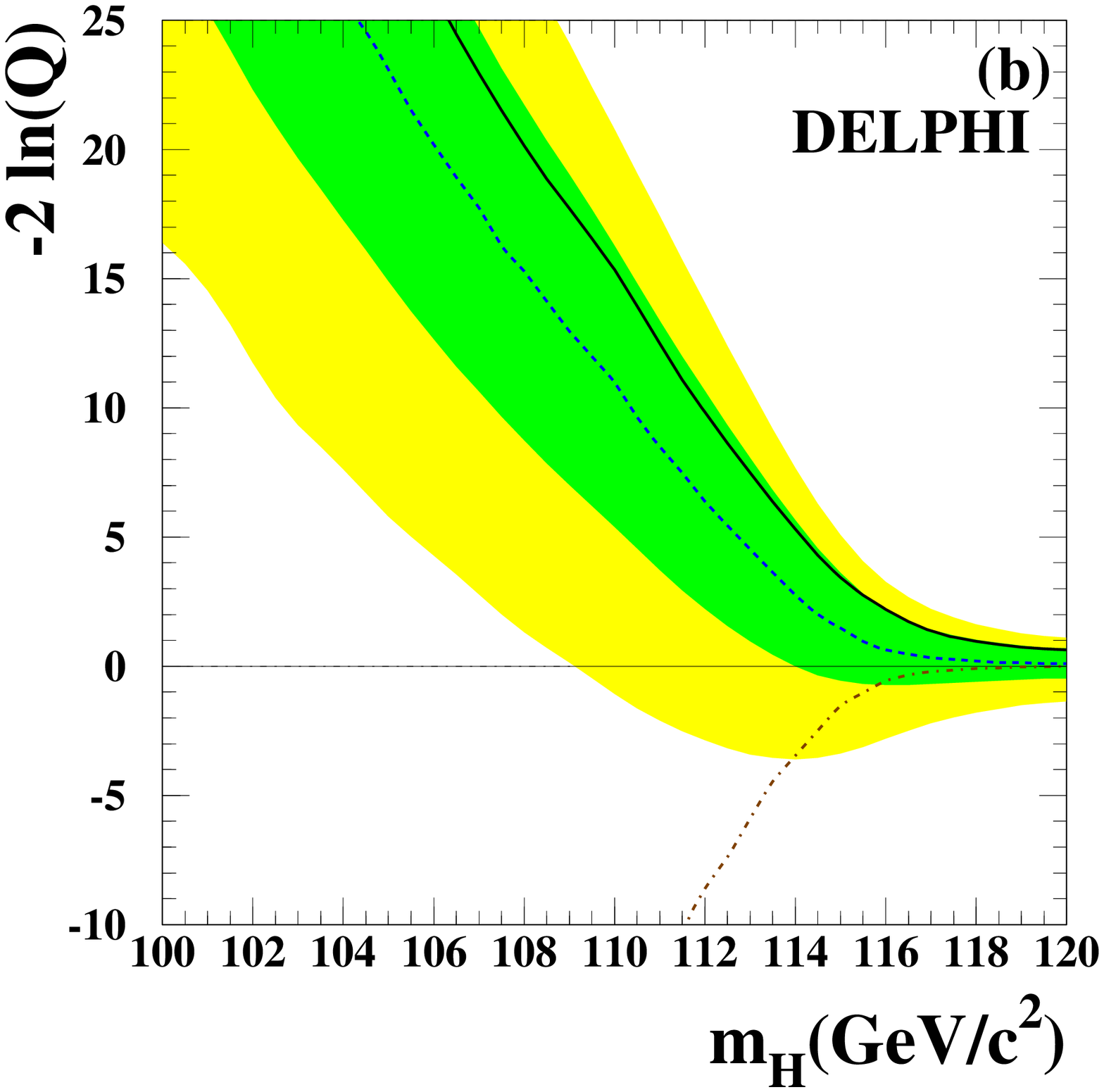,width=0.40\textwidth}\\
\epsfig{figure=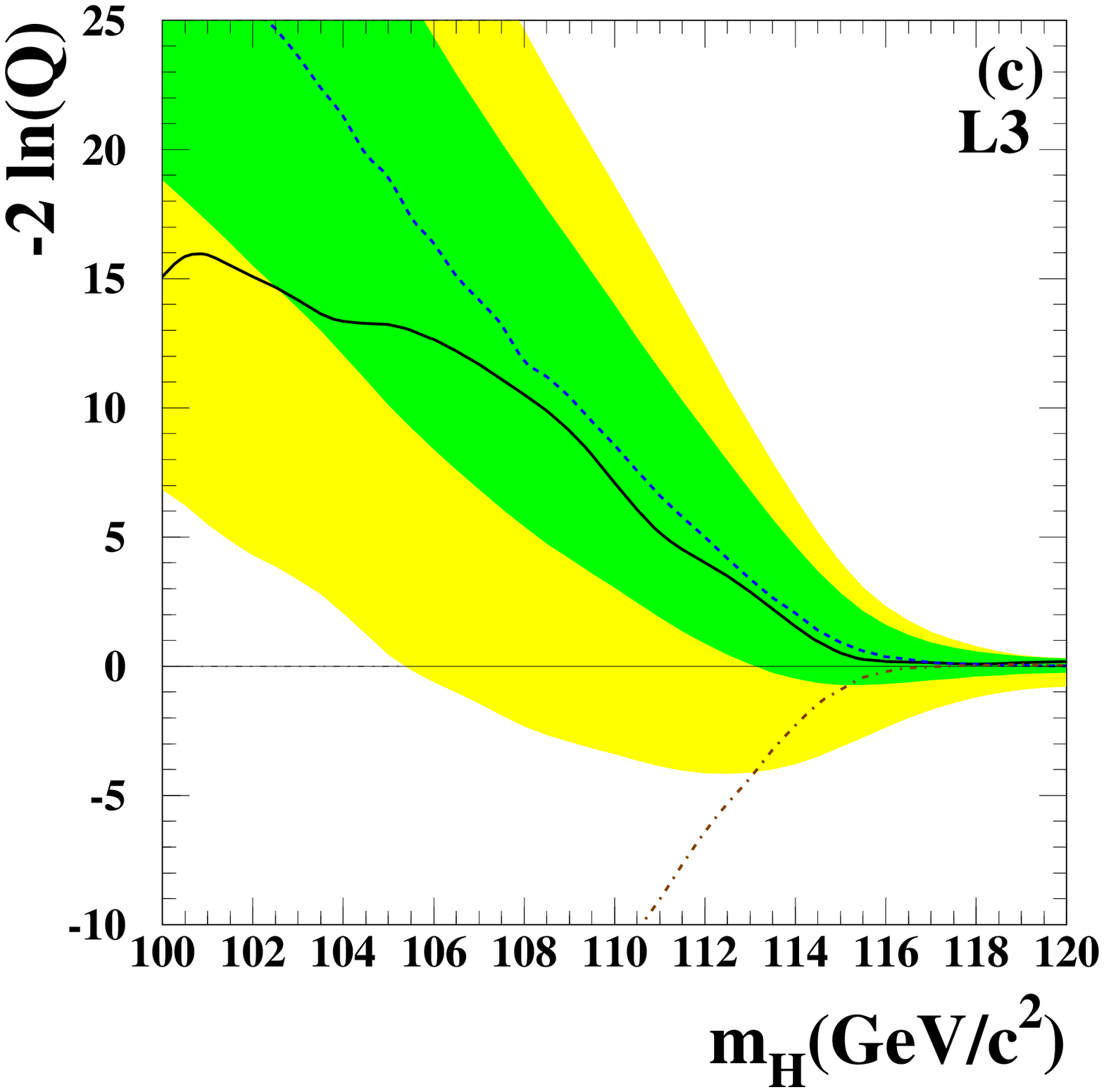,width=0.40\textwidth}
\epsfig{figure=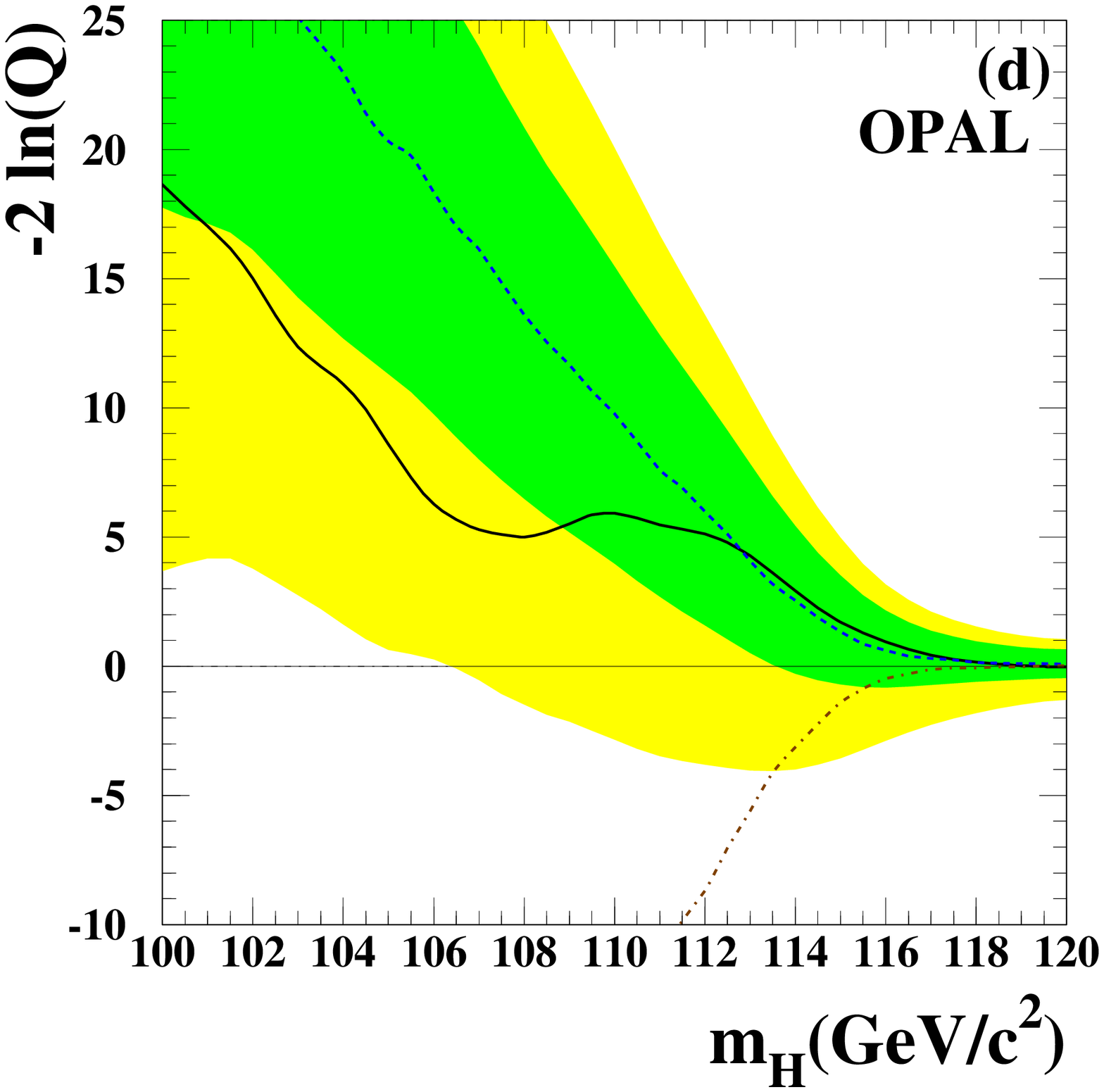,width=0.40\textwidth}\\
\epsfig{figure=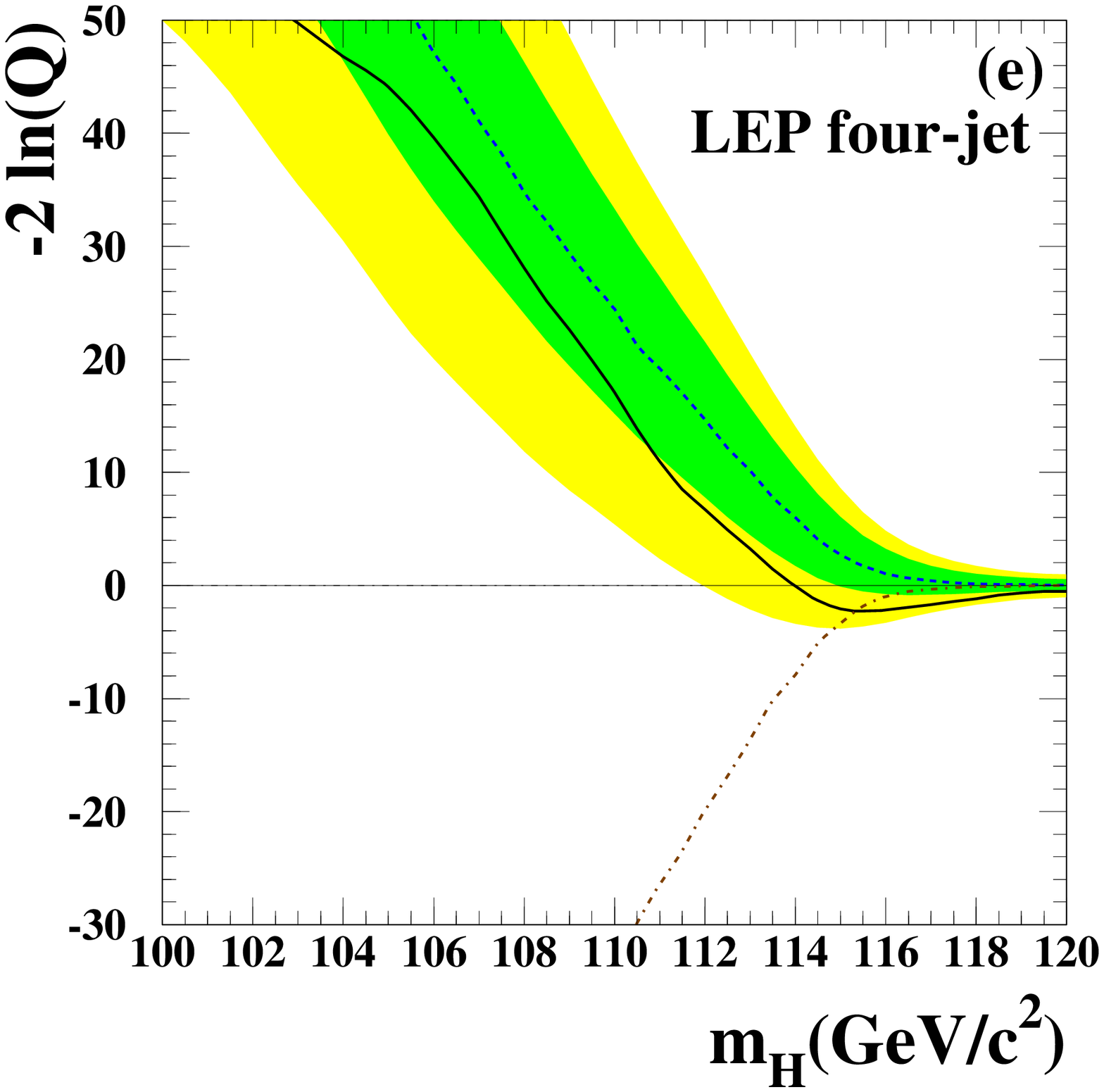,width=0.40\textwidth}
\epsfig{figure=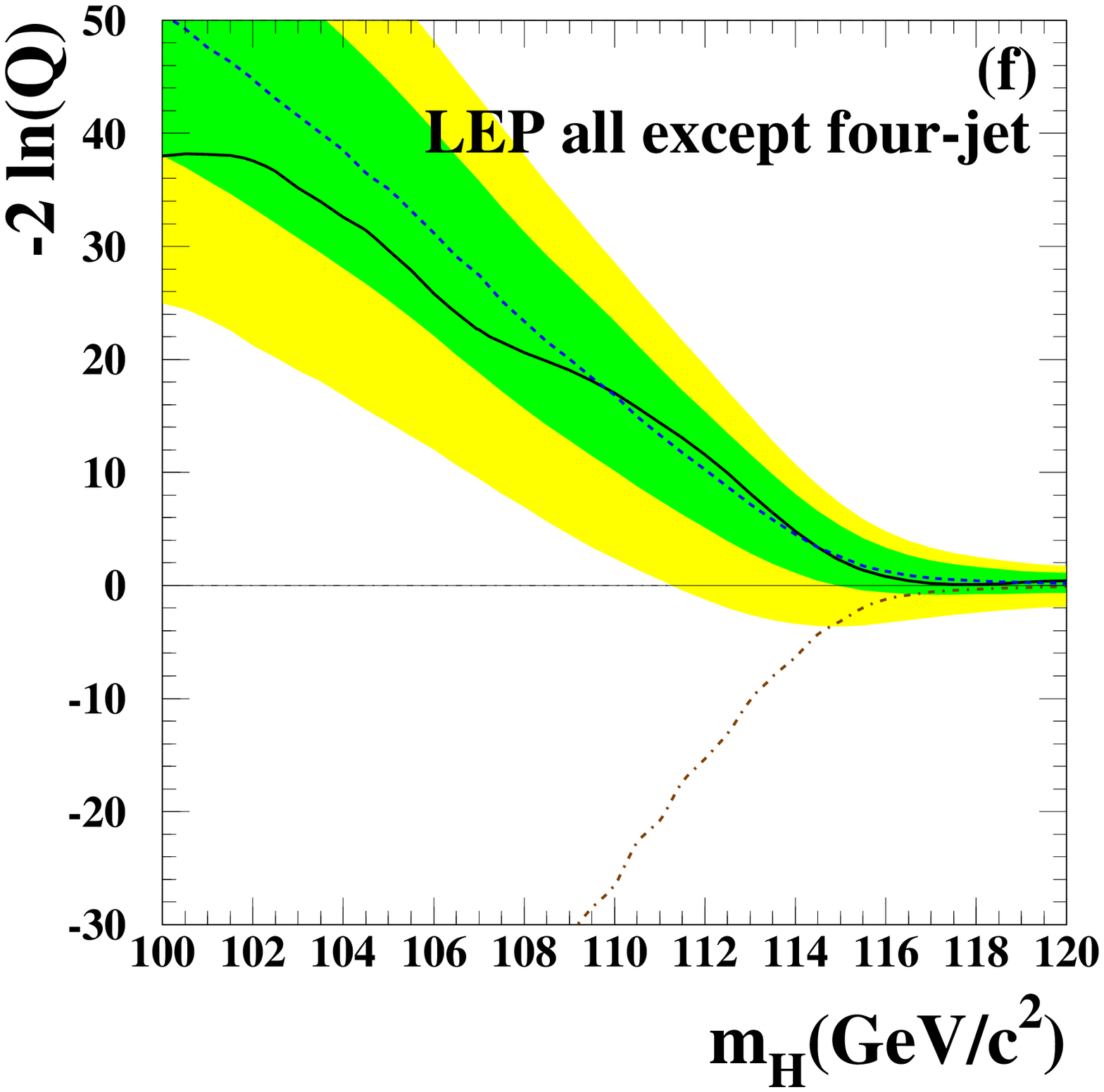,width=0.40\textwidth}\\
\caption[]{\small Observed and expected behaviour of the test statistic $-2\ln Q$ as a function of the test mass
$\mH $ when the combination procedure is applied to subsets of the LEP data. Plots (a) to (d): data sets from individual experiments; 
(e): the four-jet final state and (f): all except the four-jet final state,  
with the data of the four experiments combined. The same notation as in Figure~\ref{fig:adlo-lnq} is used.
\label{fig:a-d-l-o-lnq}}
\end{center}
\end{figure}
\begin{figure}[htb]
\begin{center}
\epsfig{figure=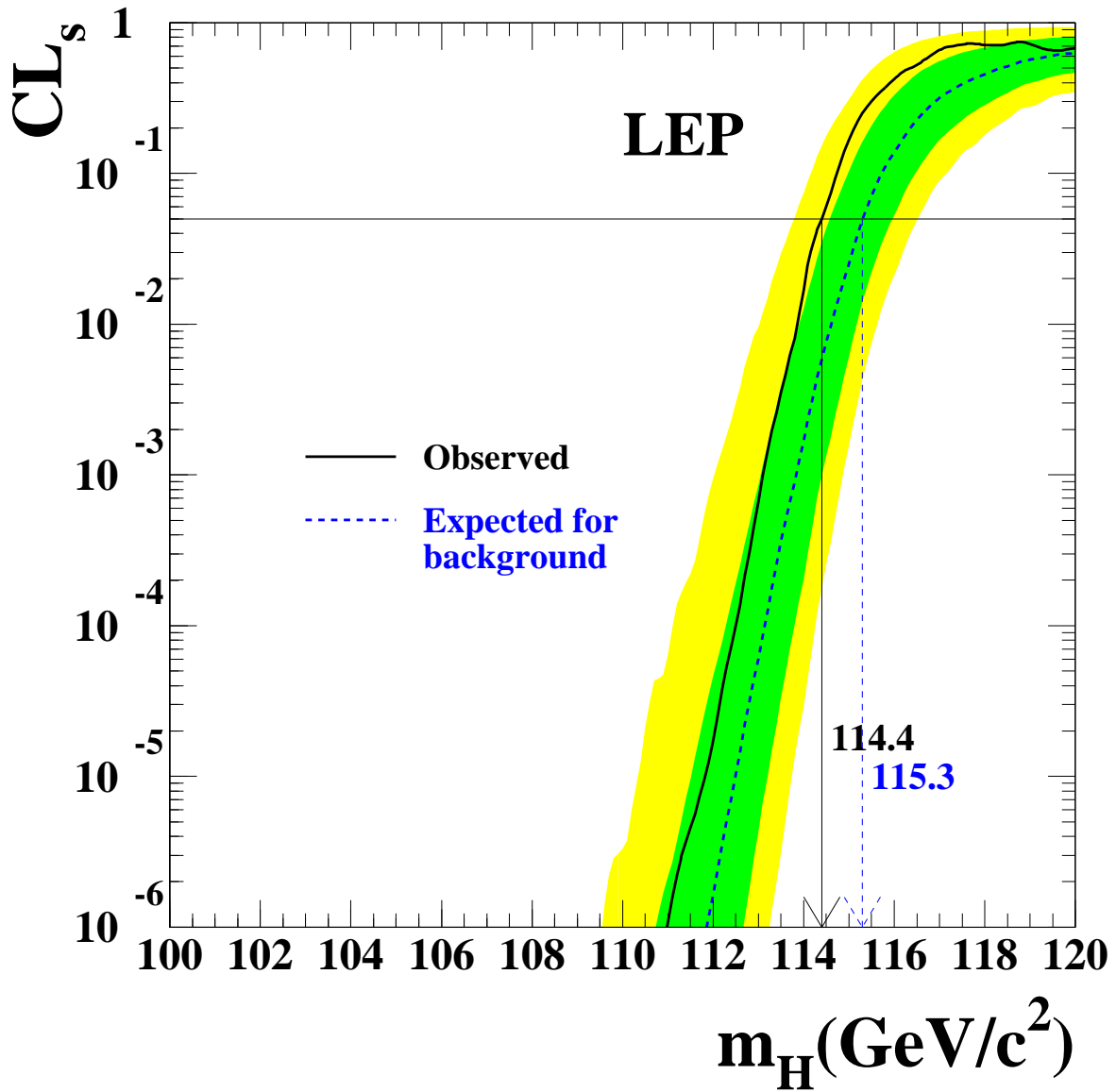,width=\textwidth}
\caption[]{\small The ratio $\cls=\clsb/\clb$ for the signal plus background hypothesis. Solid line: observation; dashed line:
median background expectation. The dark and light shaded bands around the median
expected line correspond to the 68\% and 95\% probability bands. The intersection of the horizontal line 
for $\cls = 0.05$ with the observed curve is used to define the 95\%
confidence level
lower bound on the mass of the Standard Model Higgs boson.
\label{fig:adlo-cls}}
\end{center}
\end{figure}

\begin{thebibliography}{99}
\bibitem{lhwg}
The LEP  Working Group for Higgs Boson Searches consists of members of the four LEP 
collaborations and of theorists among whom S. Heinemeyer and G. Weiglein 
are authors of this paper.
\end{thebibliography}
\end{document}